\documentclass[lettersize,journal]{IEEEtran}
\usepackage{amsmath,amsfonts}
\usepackage{algorithmic}
\usepackage{algorithm}
\usepackage{array}
\usepackage[caption=false,font=normalsize,labelfont=sf,textfont=sf]{subfig}
\usepackage{textcomp}
\usepackage{stfloats}
\usepackage{url}
\usepackage{verbatim}
\usepackage{graphicx}
\usepackage{cite}
\usepackage{tabu}                      
\usepackage{booktabs}                  
\usepackage{lipsum}                    
\usepackage{mwe}                       
\usepackage{color}
\usepackage{xcolor}
\usepackage{mathptmx}                  
\usepackage{amssymb}
\usepackage{multirow}
\usepackage{utfsym}
\usepackage{orcidlink}
\hypersetup{colorlinks=true,linkcolor=black,citecolor=black,urlcolor=black}
\usepackage{svg}
\newcommand{\linkedsubsection}[2]{%
  \subsection{\texorpdfstring{\protect\hyperlink{#1}{#2}}{#2}}%
}

\begin{document}

\title{Conflict Resolution Strategies for Co-manipulation of Virtual Objects Under Non-disjoint Conditions}

\author{Xian Wang$^{\orcidlink{0000-0003-1023-636X}}$,~\IEEEmembership{Student Member,~IEEE,} Xuanru Cheng$^{\orcidlink{0009-0003-9470-5537}}$,~\IEEEmembership{Student Member,~IEEE,} Rongkai Shi$^{\orcidlink{0000-0001-8845-6034}}$,~\IEEEmembership{} Lei Chen$^{\orcidlink{0000-0002-1524-3000}}$,~\IEEEmembership{} Jingyao Zheng$^{\orcidlink{0000-0001-8920-308X}}$,~\IEEEmembership{Student Member,~IEEE,} Hai-Ning Liang$^{\orcidlink{0000-0003-3600-8955}}$,~\IEEEmembership{Member,~IEEE,} Lik-Hang Lee$^{\orcidlink{0000-0003-1361-1612}}$,~\IEEEmembership{Senior Member,~IEEE,}
\thanks{Manuscript received April 19, 2021; revised August 16, 2021. This research was supported by the Hong Kong Polytechnic University's Start-up Fund for New Recruits (No. P0046056), Departmental General Research Fund (DGRF) from HK PolyU ISE (No. P0056354), and PolyU RIAM -- Research Institute for Advanced Manufacturing (No. P0056767). Xian Wang and Jingyao Zheng were supported by a grant from the PolyU Research Committee under student account codes RMHD and RMCU, respectively. \textit{(Corresponding author: Lik-Hang Lee.)}}
\thanks{This work involved human subjects or animals in its research. The experimental procedures and protocols was approved by the Hong Kong Polytechnic University ethics review board (Application No. HSEARS20250724002).}
\thanks{Xian Wang, Xuanru Cheng, Jingyao Zheng and Lik-Hang Lee are with the Department of Industrial and Systems Engineering, the Hong Kong Polytechnic University, Hong Kong (e-mail: \{xiann.wang, jiabao.cheng, jingyao.zheng\}@connect.polyu.hk; \textit{lik-hang.lee@polyu.edu.hk}). Rongkai Shi and Hai-Ning Liang is with the Hong Kong University of Science and Technology (Guangzhou) (e-mail: \{rongkaishi, hainingliang\}@hkust-gz.edu.cn). Lei Chen is with Hebei GEO University (e-mail: chenlei@hgu.edu.cn).}
}
\markboth{Journal of \LaTeX\ Class Files,~Vol.~14, No.~8, August~2021}%
{Wang \MakeLowercase{\textit{et al.}}: Conflict Resolution Strategies for Co-manipulation}

\IEEEpubid{0000--0000/00\$00.00~\copyright~2021 IEEE}
\maketitle
\begin{abstract}
  Virtual Reality (VR) co-manipulation enables multiple users to collaboratively interact with shared virtual objects. However, existing research treats objects as monolithic entities, overlooking scenarios where users need to manipulate different sub-components simultaneously. This work addresses conflict resolution when users select overlapping vertices (non-disjoint sets) during co-manipulation. We present a comprehensive framework comprising preventive strategies (Object-level and Action-level Restrictions) and reactive strategies (computational conflict resolution). Through two user studies with 76 participants (38 pairs), we evaluated these approaches in collaborative wireframe editing tasks. Study 1 identified Averaging as the optimal computational method, balancing task efficiency with user experience. Study 2 highlighted that Action-level Restriction, which permits overlapping selections but restricts concurrent identical operations, achieved better performance compared to exclusive object locking. Reactive strategies using averaging provided smooth collaboration for experienced users, while second-user priority enabled quick corrections. Our findings indicate that optimal strategy selection depends on task requirements, user expertise, and collaboration patterns. Based on the findings, we provide design implications for developing VR collaboration systems that support flexible sub-components manipulation while maintaining collaborative awareness and minimizing conflicts.
\end{abstract}


\begin{figure*}[htb]
  \centering
  \includegraphics[width=\linewidth]{Figures/Teaser.pdf}
  \caption{Overview of conflict resolution strategies for NDS conditions in VR. (a) \textit{Object-level Restriction} enforces exclusive access control where users can only manipulate their own vertex groups (User 1: \textcolor[RGB]{255,59,47}{red}, User 2: \textcolor[RGB]{0,121,255}{blue}) with no overlapping selections permitted. (b) \textit{Action-level Restriction} allows overlapping vertex selections (\textcolor[RGB]{175, 82, 221}{purple} indicates joint vertices) but prevents concurrent identical operations on overlapping points, users must perform different transformation types when manipulating joint vertices. The system applies both operations to the joint vertices simultaneously. (c)\textit{Reactive Strategies} permit both overlapping selections and concurrent identical operations without restrictions; conflicts are resolved computationally using \textit{Additive Combination}, \textit{Average}, or \textit{Intersection} methods when users simultaneously perform the same operation on joint vertices. For strategies (2) and (3), \includegraphics[width=8pt]{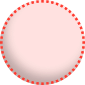} and \includegraphics[width=8pt]{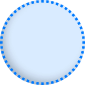} represent User 1 and User 2's respective individual actions, while \includegraphics[width=8pt]{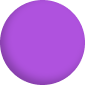} results show what both users see in real-time.}
  \label{fig:teaser}
\end{figure*}


\section{Introduction}
\IEEEPARstart{T}HE metaverse emphasises a concept of virtual social space \cite{8ca911d754f448fdb959f34677db6f95} and allows multiple users to collaborate on tasks of shared interest~\cite{benford2001collaborative}, which are regarded as collaborative virtual environments (CVEs). Diversified CVEs exist, including design~\cite{Sereno2022Collaborative,Li2022ARCritique}, training~\cite{Schild2018Applying,Chheang2019Collaborative}, and education~\cite{Laine2024Collaborative,Radu2023How}. In CVEs, users collaboratively manipulate virtual objects, share spatial awareness, and perform collaboration tasks as if they were co-located~\cite{Soares2018EGO-EXO,Grandi2019Characterizing}. For example, users may jointly manipulate a 3D object by resizing, rotating, or modifying its structure~\cite{Chen2021On,Hubenschmid2023Colibri}, or they can collectively annotate and rearrange virtual learning materials~\cite{Yang2022Towards}. This refers to co-manipulation among users (also known as collaborative or cooperative manipulation), where multiple users interact simultaneously with virtual object(s)~\cite{Grandi2019Characterizing,Pinho2002Cooperative}. 


However, existing co-manipulation research predominantly treats virtual objects as monolithic, indivisible entities, overlooking a key characteristic of objects: they typically consist of multiple interconnected sub-components that can be manipulated independently. whereas many real-world collaborative tasks require fine-grained co-manipulation. 
For instance, a virtual car consists of numerous distinct parts, such as the engine block, doors, wheels, and seats. During assembly, each part requires precise positioning and orientation while maintaining critical spatial relationships with the others. Similarly, a piece of furniture might consist of drawers, doors, and shelves that users need to adjust individually, or a mechanical assembly might contain numerous parts requiring independent positioning. This characterization is particularly evident in virtual content, where 3D models are composed of thousands of vertices, edges, and faces that designers manipulate selectively. 
In virtual reality (VR) painting applications like Tilt Brush~\cite{tiltbrush}, artists work with multiple brush strokes and layers simultaneously, while in modeling tools like Gravity Sketch~\cite{gravitysketch}, designers frequently select and transform collections of vertices to reshape their creations. 
\IEEEpubidadjcol
We consider a fundamental case of two users (A \& B) working on a set of vertices in a virtual object: $V=\{v_1,v_2,v_3,v_4\}$. When two users, who own different interests in a virtual object, select overlapping sets of the object's sub-components, complex conflicts arise. We further describe this user interaction in terms of a collection of sub-objects, utilizing a mathematical concept known as set theory: the non-disjoint set (NDS). At a particular time, NDS refers to the users' co-manipulation (or selection) of at least one common vertex, whereas the remaining vertices are selected separately by each user. 
Prior work primarily handles conflict resolution by all sub-components as an indivisible entity (i.e., $V$), thereby limiting the actions of another user if $V$ $\neq$ NDS. Various strategies have been proposed accordingly to prevent and resolve conflicts, including locking mechanisms~\cite{Chen2021On,Li2003VSculpt}, separation of degrees of freedom (DOF)
\cite{Wieland2021Separation,Grandi2018Design}, and input computations such as averaging or merging \cite{Wieland2021Separation,Soares2018EGO-EXO}. 
In contrast, our work considers the user's needs for the possibility of co-manipulation on the vertices of common interests by two users (i.e., NDS), while simultaneously maintaining each user's freedom to work on the remaining vertices. For example, User A selects vertices $\{v_1,v_2,v_3\}$ to rotate, and User B simultaneously selects vertices $\{v_2,v_3,v_4\}$ to scale. 
Thus, a system should resolve conflicting operations on the joint vertices $\{v_2,v_3\}$ while preserving non-conflicting transformations on the disjoint vertices $\{v_1\}$ and $\{v_4\}$. Such co-manipulation remains a crucial yet underexplored problem in virtual social space. 

\hypertarget{RQ}{}
To address this gap, we propose a framework for conflict resolution to accommodate the complexity of co-manipulation scenarios involving joint and disjoint vertices in immersive environments (Fig.~\ref{fig:teaser}). We group conflict-resolution strategies into \textbf{(i)} restriction-based strategies that prevent conflicts by restricting operations when selections overlap and \textbf{(ii)} computation-based strategies that resolve conflicts by combining users’ inputs on shared vertices; we refer to these as \textit{preventive} and \textit{reactive strategies}, respectively. We implement this framework in a wireframe-editing system and evaluate it in two controlled user studies designed to elicit NDS. We investigate the following research questions:

\begin{itemize}
    \item[\textbf{RQ1:}] How can conflict resolution strategies in immersive environments be extended from whole-object manipulation to overlapping vertex sets?
    \item[\textbf{RQ2:}] When users concurrently manipulate the vertices, how do different computational methods for combining their inputs affect task performance and user experience?
    \item[\textbf{RQ3:}] How do preventive and reactive strategies differentially affect user behavior, performance, and experience in multi-object co-manipulation?
\end{itemize}
\section{Related Work}

\subsection{Conflict Resolution Strategies in Co-manipulation}
\label{Sec:realted_work_Strategies}
In CVEs, conflicts naturally arise when co-manipulators have different intentions or attempt to control the same virtual object simultaneously. In addition to the basic locking mechanism~\cite{Li2003VSculpt}, recent literature has explored various conflict resolution techniques. In the following, we summarize the key approaches and challenges of these solutions.

\paragraph{Enhancing User Awareness} These solutions emphasize the enhancement of mutual awareness between collaborators through various methods, such as adjusting user perspectives or employing different visualization designs to mitigate conflicts. For example, \textit{EGO-EXO} provides users with asymmetric viewpoints (egocentric vs. exocentric), allowing them to perceive different aspects of the task and thus reduce unintentional interference~\cite{Soares2018EGO-EXO}, Wang et al. assign a dominant manipulator based on collaborators’ viewpoint quality~\cite{wang2021vr}, and Liu et al. extend this with a manipulation guidance field that steers users to complementary viewpoints and roles~\cite{liu2024manipulation}. Furthermore, multi-user interactions in immersive environments benefit from perceptual cues. Both \textit{Blended Whiteboard} and Khan et al. demonstrate that highlighting personal versus shared objects can alleviate accidental ``blocking'' or occlusion, thus improving collaborators’ understanding of one another’s virtual items and reducing the likelihood of conflict~\cite{Gronbaek2024Blended,Khan2024Dont}. Similarly, studies by Ouedraogo et al. and Park show that clarifying workspace boundaries and ownership (e.g., through color-coding or partitioning) can effectively prevent or resolve conflicts~\cite{Ouedraogo2024Where,Park2020ARLooper}. In addition, Baron uses constrained handles and visual feedback to coordinate joint operations~\cite{baron2016collaborativeconstraint}. Haptic feedback can also improve awareness. Auda et al. used haptic props to indicate object ownership, allowing users to feel when a collaborator attempts to move the object simultaneously~\cite{Auda2021Im}.

\paragraph{Separating or Composing User Actions} These solutions can be categorized into three main approaches: 1) DOF separation, 2) Composition (merging transformations), and 3) Hybrid approaches~\cite{Wieland2021Separation}. In the Separation approach, each user exclusively controls a subset of the object's six DOF, such as one user managing translations while the other handles rotations~\cite{Grandi2018Design,Pinho2002Cooperative,Grandi2019Characterizing,Soares2018EGO-EXO}. Alternatively, users may control specific rotation axes separately. By assigning each DOF to one particular user and applying the inputs independently, the system ensures that no single DOF receives conflicting commands. In Composition, multiple users can simultaneously manipulate the same DOFs (e.g., both can push, pull, or rotate the object at once). The system then blends their input into a unified transformation. Various research prototypes define different merge policies to mathematically combine these inputs, such as \textit{summation}~\cite{Hubenschmid2023Colibri}, \textit{averaging}~\cite{Duval2006SkeweR}, \textit{weighted combination} (i.e., each user is assigned a weight based on summation)~\cite{Kai2006The}, or \textit{common component} (i.e., calculate only the overlap of transformations)~\cite{Ruddle2002Symmetric}. Hybrid approaches enable dynamic switching between Separation and Composition. For example, if one user requires precise control, the system may temporarily restrict the influence of other users on certain DOFs~\cite{Wieland2021Separation}.

\paragraph{Versioning and Rollback} When conflicts escalate, undo mechanisms or branching can help. Rasch et al. highlighted the importance of undo in mitigating collaboration conflicts~\cite{Rasch2024Just}. Similarly, \textit{VRGit}~\cite{Zhang2023VRGit} introduced a Git-like version control system for VR content, enabling teams to restore or merge edits and thereby reduce conflict-related issues. \textit{Photoportals}~\cite{Kunert2014Photoportals} allow users to manipulate alternative ``portals'', reducing conflicts in the main scene. \textit{Spacetime}~\cite{Xia2018Spacetime} introduced the concept of parallel editing by groups, with the final version decided by discussion. Similarly, \textit{CIDER}~\cite{Pintani2023CIDER} designed private layers for asynchronous editing and submission, where collaborators propose changes privately and merge them into the shared final version only after resolving conflicts.


\subsection{Synthesis of Prior Work and Design Opportunities}
\label{sec:Design_Opportunities}
Building upon the foundation of existing co-manipulation research, we extend conflict resolution strategies in an immersive space shared by two users. In this section, we first synthesize the design space based on previous work to identify key design opportunities. Accordingly, we present our conflict resolution framework specifically tailored for co-manipulation in VR.

\begin{table}[htbp]
\centering
\caption{Feature Comparison and Categorization of Co-manipulation Object Systems}
\setlength{\tabcolsep}{5pt}
\resizebox{\linewidth}{!}{ 
\begin{tabular}{c|ccc|ccc|ccc|cc|cccc}
\toprule
Key Literature & \multicolumn{3}{c}{A} & \multicolumn{3}{c}{B} & \multicolumn{3}{c}{C}  & \multicolumn{2}{c}{D} & \multicolumn{4}{c}{E}\\
\midrule
 & 
\rotatebox{90}{{HMD VR}} & 
\rotatebox{90}{{Handheld AR}} & 
\rotatebox{90}{{2D Screen}} & 
\rotatebox{90}{{Object Restriction}} & 
\rotatebox{90}{{Action Restriction}} & 
\rotatebox{90}{{Reactive Strategies}} & 
\rotatebox{90}{{Translation}} & 
\rotatebox{90}{{Rotation}} & 
\rotatebox{90}{{Scaling}} & 
\rotatebox{90}{{Sub-component}} & 
\rotatebox{90}{{Monolithic Entities}} & 
\rotatebox{90}{{Additive Combination}} & 
\rotatebox{90}{{Averaging}} & 
\rotatebox{90}{{Intersection}} & 
\rotatebox{90}{{Other}} \\
\midrule
Wieland et al.\cite{Wieland2021Separation} & & $\checkmark$ & & & $\checkmark$ & $\checkmark$ & $\checkmark$ & $\checkmark$ & & & $\checkmark$ & $\checkmark$ & & & $\checkmark$ \\
Grandi et al.\cite{Grandi2018Design} & & $\checkmark$ & & $\checkmark$ & $\checkmark$ & $\checkmark$ & $\checkmark$ & $\checkmark$ & $\checkmark$ & & $\checkmark$ & $\checkmark$ & & $\checkmark$ & $\checkmark$ \\
Grandi et al.\cite{Grandi2019Characterizing} & $\checkmark$ & $\checkmark$ & & $\checkmark$ & $\checkmark$ & $\checkmark$ & $\checkmark$ & $\checkmark$ & $\checkmark$ & & $\checkmark$ & $\checkmark$ & & &  \\
Pinho et al.\cite{pinho2008cooperative} & $\checkmark$ & & & & $\checkmark$ & & $\checkmark$  & $\checkmark$  & & & $\checkmark$  & & & & \\
Soares et al.\cite{Soares2018EGO-EXO} & $\checkmark$ & & & & $\checkmark$ & & $\checkmark$ & $\checkmark$ & & & $\checkmark$ & & & & \\
Pinho et al.\cite{Pinho2002Cooperative} & $\checkmark$ & & & & $\checkmark$ & & $\checkmark$ & $\checkmark$ & & & $\checkmark$ & & & & \\
Ruddle et al.\cite{Ruddle2002Symmetric} & & & $\checkmark$ & & & $\checkmark$ & $\checkmark$ & $\checkmark$ & & & $\checkmark$ & & $\checkmark$ & $\checkmark$ & \\
Duval et al.\cite{Duval2006SkeweR} & & & $\checkmark$ & & & $\checkmark$ & $\checkmark$ & $\checkmark$ & & & $\checkmark$ & & & & $\checkmark$ \\
Giege et al.\cite{Riege2006bent} & & & $\checkmark$ & & & $\checkmark$ & $\checkmark$ & $\checkmark$ & & & $\checkmark$ & & $\checkmark$ & & \\
Grandi et al.\cite{Grandi2017Design} & & & $\checkmark$ & & & $\checkmark$ & $\checkmark$ & $\checkmark$ & $\checkmark$ & & $\checkmark$ & $\checkmark$ & $\checkmark$ & & $\checkmark$ \\
Chenechal et al.\cite{Chenechal2016Giant} & $\checkmark$ & & & & $\checkmark$ & $\checkmark$ & & $\checkmark$ & $\checkmark$ & & $\checkmark$ & $\checkmark$ & & & \\
\midrule
Ours & $\checkmark$ & & & $\checkmark$ & $\checkmark$ & $\checkmark$ & $\checkmark$ & $\checkmark$ & $\checkmark$ & $\checkmark$ & & $\checkmark$ & $\checkmark$ & $\checkmark$ & \\
\bottomrule
\end{tabular}}
\label{tab:comparison}
\end{table}

Our analysis of existing co-manipulation literature (Tab.~\ref{tab:comparison}) focuses specifically on object manipulation systems. To make this synthesis explicit and reproducible, we compiled Tab.~\ref{tab:comparison} via a structured feature-coding of prior systems along five dimensions: (A) display technology, (B) conflict-resolution strategy type, (C) transformation support, (D) object treatment granularity, and (E) reactive strategy calculation. For each paper, we extracted whether the system explicitly supported each feature and marked it accordingly. This coding reveals several under-covered areas that motivate our problem setting. \textbf{First}, Prior work has focused only on monolithic, indivisible entities rather than manipulable sub-components (e.g., vertices/parts), leaving limited support for modern VR tasks where users need to co-edit multiple interconnected elements (e.g., vertex groups in 3D modeling or collections in design tasks). \textbf{Second}, many systems provide incomplete transformation coverage, commonly emphasizing translation/rotation while omitting scaling or unified multi-DOF support~\cite{Wieland2021Separation,pinho2008cooperative,Pinho2002Cooperative,Ruddle2002Symmetric,Duval2006SkeweR,Riege2006bent}. \textbf{Third}, although multiple computational policies for composing concurrent actions have been proposed (e.g., additive combinations, averaging, intersection), they are rarely evaluated side-by-side under a harmonized task and platform, making it difficult to compare their suitability for NDS. \textbf{Finally}, deployment constraints remain salient: solutions on 2D screens~\cite{Ruddle2002Symmetric} or handheld AR devices~\cite{Wieland2021Separation,Grandi2018Design} inherently limit DOF bandwidth and often encourage DOF constraints, whereas immersive VR enables unrestricted 6-DOF manipulation and thus introduces new conflict patterns that these systems do not fully address. 
According to Table~\ref{tab:comparison}, we derive design opportunities from the broader literature discussed earlier in Sec.~\ref{Sec:realted_work_Strategies}. Enhanced user awareness strategies are non-conflicting and can be integrated into our framework, where we employ color coding to provide visual feedback. However, we exclude version control and rollback mechanisms as they introduce delays that are incompatible with real-time collaborative operations~\cite{Rasch2024Just,Zhang2023VRGit,Xia2018Spacetime}.
In sum, we conclude with five design opportunities: 
\begin{itemize}
    \item[\textbf{(1)}] Supporting unrestricted \textbf{multiple elements manipulation} in real-time;
    \item[\textbf{(2)}] Implementing \textbf{complete transformation} operations in NDS scenarios;
    \item[\textbf{(3)}] Evaluating multiple conflict resolution strategies and computational approaches in a \textbf{unified test environment};
    \item[\textbf{(4)}] Operating in immersive VR \textbf{without DOF constraints}; and
    \item[\textbf{(5)}] Incorporating awareness-enhancing \textbf{visual cues}.
\end{itemize}


\begin{table*}[htbp]
\caption{Examples for Conflict Resolution Framework}
\resizebox{\linewidth}{!}{
\begin{tabular}{cl|p{5pt}p{5pt}|p{130pt}p{170pt}|ll}
\toprule
\multicolumn{2}{c|}{Strategy} & A & B & \multicolumn{1}{c}{User 1} & \multicolumn{1}{c|}{User 2} & \multicolumn{2}{c}{Result} \\
\midrule
\multirow{4}{*}{\rotatebox{90}{Preventive}}  & \multirow{2}{*}{Object-level Restriction}  & \multirow{2}{*}{$\usym{2613}$} & \multirow{2}{*}{$\usym{2613}$} & select $\{\textcolor[RGB]{255,59,47}{v_1, v_2, v_3, v_4}\}$ $\rightarrow$ locked & ($\usym{2613}$) select $\{v_3, v_4, v_5, v_6\}$ $\rightarrow$ $\{v_3, v_4\}$ prevented & \multicolumn{2}{l}{only disjoint vertices, no conflicts occur}\\
                                        &                           &  &  &  & ($\checkmark$) select $\{\textcolor[RGB]{0,121,255}{v_5, v_6}\}$ & & \\
                                        & \multirow{2}{*}{Action-level Restrictions} & \multirow{2}{*}{$\checkmark$} & \multirow{2}{*}{$\usym{2613}$} & select $\{\textcolor[RGB]{255,59,47}{v_1, v_2,}\textcolor[RGB]{175, 82, 221}{v_3, v_4}\}$ rotation & ($\usym{2613}$) select $\{v_3, v_4, v_5, v_6\}$ rotation & \multicolumn{2}{l}{$\{\textcolor[RGB]{255,59,47}{v_1, v_2}\}$ rotated, $\{\textcolor[RGB]{0,121,255}{v_5, v_6}\}$ scaled}\\ 
                                        &                           &  &  &  & ($\checkmark$) select $\{\textcolor[RGB]{175, 82, 221}{v_3, v_4, }\textcolor[RGB]{0,121,255}{v_5, v_6}\}$ scaling/translation & \multicolumn{2}{l}{$\{\textcolor[RGB]{175, 82, 221}{v_3, v_4}\}$ rotated \& scaled/translated}\\
\midrule
\multirow{4}{*}{\rotatebox{90}{Reactive}}    & Additive Combination      & $\checkmark$ & $\checkmark$ & \multirow{4}{*}{translate $\{\textcolor[RGB]{255,59,47}{v_1, v_2,}\textcolor[RGB]{175, 82, 221}{v_3, v_4}\}$ by (-0.2,0.1,0)} & \multirow{4}{*}{translate $\{\textcolor[RGB]{175, 82, 221}{v_3, v_4,}\textcolor[RGB]{0,121,255}{v_5, v_6}\}$ by (0.4,0.2,0.1)} & $\{\textcolor[RGB]{175, 82, 221}{v_3, v_4}\}$ $\rightarrow$ (0.2,0.3,0.1) & \multirow{2}{*}{$\{\textcolor[RGB]{255,59,47}{v_1, v_2}\}$ $\rightarrow$ (-0.2,0.1,0)}\\
                                        & Averaging                 & $\checkmark$ & $\checkmark$ & & & $\{\textcolor[RGB]{175, 82, 221}{v_3, v_4}\}$ $\rightarrow$ (0.1,0.15,0.05) & \\
                                        & Intersection              & $\checkmark$ & $\checkmark$ & & & $\{\textcolor[RGB]{175, 82, 221}{v_3, v_4}\}$ $\rightarrow$ (0,0.1,0.1) & \multirow{2}{*}{$\{\textcolor[RGB]{0,121,255}{v_5, v_6}\}$ $\rightarrow$ (0.4,0.2,0.1)}\\
                                        & Second-user priority      & $\checkmark$ & $\checkmark$ & & & $\{\textcolor[RGB]{175, 82, 221}{v_3, v_4}\}$ $\rightarrow$ (0.4,0.2,0.1)& \\
\bottomrule
\end{tabular}}
{\scriptsize  \textit{(A) Overlap Vertices Groups Support, (B) Same Transformation Support under Overlap Vertices Groups; Virtual object $V=\{v_1,v_2,v_3,v_4,v_5,v_6,v_7,v_8\}$; \textcolor[RGB]{255,59,47}{Red vertices} $\rightarrow$ disjoint vertices by User 1, \textcolor[RGB]{0,121,255}{Blue vertices} $\rightarrow$ disjoint vertices by User 2, \textcolor[RGB]{175, 82, 221}{Purple vertices} $\rightarrow$ joint vertices; User 1 conducts the operation first.}}
\label{tab:Framework_example}
\end{table*}
 
\section{Conflict Resolution Framework and System Implementation}
To address \textbf{RQ1}, we developed a comprehensive conflict resolution framework. Based on the identified design opportunities, we propose solution strategies to mitigate conflicts in NDS scenarios and design a unified VR test environment for evaluation that supports multiple users in completing model editing tasks.

\subsection{Supporting Unrestricted Multiple Elements Manipulation in Real-time}

To evaluate conflict resolution strategies in NDS co-manipulation scenarios, we developed a wireframe-based model editing system in VR that supports multi-user interaction. The system can generate wireframe models with varying vertex counts and geometric complexity to test collaborative manipulation under diverse conditions. A client--server architecture synchronizes all manipulation operations in real time across users (see Fig.~\ref{fig:VE}(c) for the physical setup of the workbench). To support spatial awareness, each user sees the collaborator as a simple green avatar consisting of a head and two hands. The transformation and collaboration operations supported by the system are described below.

\subsection{Implementing Complete Transformation Operations in NDS scenarios}

\begin{figure}[htb]
  \centering
  \includegraphics[width=1\linewidth]{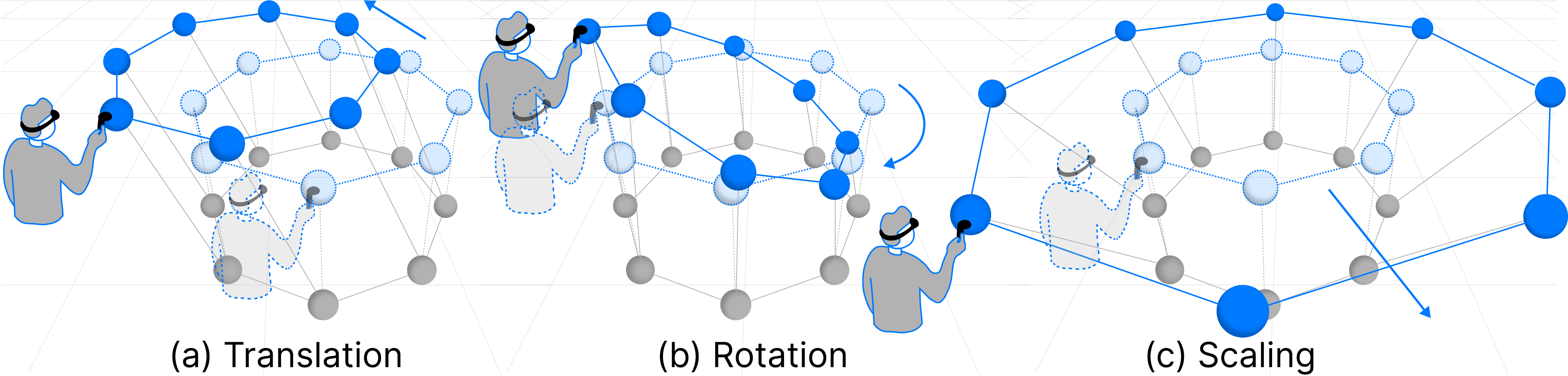}
  \caption{Multi-object transformation operations: (a) translation, (b) rotation, and (c) scaling performed relative to the group centroid.}
  \label{fig:function}
\end{figure}

The system supports three fundamental transformation operations: translation, rotation, and scaling, all performed relative to the group's centroid (Fig.~\ref{fig:function}). Users select the desired transformation through a virtual panel positioned in front of the left controller, enabling seamless switching between operation types during manipulation.

\paragraph{Group} Users can select multiple vertices $\{v_1, v_2,..., v_n\}$ using VR controllers by pressing the ``trigger'' button and form them into a logical group representing edges, faces, or arbitrary geometric structures. To create a group, users press the ``A'' button on the controller after selecting the desired vertices; pressing ``A'' again cancels the current group and releases the selected vertices. Each user can maintain only one active group at a time. 
Upon group creation, the system generates a container object positioned at the centroid of selected vertices, calculated as $c=(1/n)\Sigma_{i=1}^nv_i$. All selected vertices are then reparented to this container, establishing a hierarchical structure that facilitates collective transformations (as shown by the group of blue vertices in Fig.~\ref{fig:function}).

\paragraph{Translation} When a user grabs any vertex $v_i$ within a group, the translation mechanism maintains the relative offset between the grabbed vertex and the group container. During manipulation, the container's position is updated according to $P_{container}(t) = P_{handle}(t) - \delta$, where $P_{handle}(t)$ represents the handle's current position and $\delta$ denotes the initial offset vector. This approach ensures that all grouped vertices translate uniformly while preserving their relative spatial relationships.

\paragraph{Rotation} The rotation mechanism computes angular transformations based on the directional change of the grabbed vertex relative to the group's center. Given an initial direction vector $d_0 = (v_i-c)/||v_i-c||$ and current direction $d(t)=(v_i(t)-c)/||v_i(t)-c||$, the system calculates the rotation quaternion $q=Quaternion.FromToRotation(d_0, d(t))$ and applies it to the entire group. This enables intuitive arc-ball-style rotation where dragging any vertex orbits the entire group around its centroid.

\paragraph{Scaling} Scaling is achieved by tracking the distance variation between the grabbed vertex and the group center. The scale factor $s(t)=||v_i(t)-c||/||v_{i0}-c||$ is applied to all vertices' local positions within the group, where $||v_{i0} - c||$ is the initial distance. To maintain geometric consistency, when the user releases the grabbed vertex, the system performs a snap-back interpolation that repositions the vertex to its scaled position within the group's local coordinate frame, ensuring planar relationships are preserved for coplanar vertex sets.

\subsection{Evaluating Multiple Conflict Resolution Strategies and Computational Approaches in a Unified Environment}

\subsubsection{Conflict Resolution Strategies}
\label{sec:strategies}
Based on previous studies, we propose two main classes of conflict resolution methods for co-manipulation scenarios involving NDS. We show examples of these methods in Table~\ref{tab:Framework_example}.

\paragraph{Preventive Strategies} These strategies aim to prevent conflicts before they occur by imposing constraints on either the manipulation targets or the available actions. We categorize them into two types: \textbf{(i) Object-level Restriction}, or exclusive locking, where the group of objects selected and manipulated by one user is temporarily unavailable to other users; \textbf{(ii) Action-level Restrictions}, this means separation of transformation types, where users are assigned specific transformation capabilities (e.g., user A handles rotations and user B handles transformations).

\paragraph{Reactive Strategies} This approach will no longer impose any restriction on the selection of the user's target objects or Action, and the system computes the conflict resolution. Based on previous literature, we adopt three computational methods. In addition, motivated by observations from User Study 1 (see Sec.~\ref{sec:Study_1}), we introduce a strategy that prioritizes the most recent user input. Together, these four strategies constitute our conflict-resolution framework (Tab.~\ref{tab:Framework_example}). \textbf{(i) Additive Combination}, where the transformations are cumulative and all user inputs are directly applied to the final result ($T_{final}=T_{user1}+T_{user2}$). \textbf{(ii) Averaging}, where all users' variation inputs are averaged to ensure moderate variation ($T_{final}=(T_{user1}+T_{user2})/2$). \textbf{(iii) Intersection}, Only transformations agreed upon by both users are applied ($T_{final}=T_{user1} \cap T_{user2}$). \textbf{(iv) Second-user priority}, a last-writer-wins rule on joint vertices, for each joint vertex, the transformation with the later timestamp, while preserving each user's transformations on disjoint vertices (see Tab.~\ref{tab:Framework_example} and Sec.~\ref{sec:Study_1_Implications}).

\subsubsection{Co-manipulation Function}
\label{sec:Co-manipulation Function}
We implemented various multi-user co-manipulation functions following the strategy outlined in Sec.~\ref{sec:strategies}.

\paragraph{Object-level Restriction} Under this preventive strategy, the system enforces exclusive access to vertex groups, preventing the occurrence of NDS conditions. When a user selects a set of vertices $\{v_1, v_2,..., v_n\}$ , the system marks these vertices as locked, making them unavailable for selection by other users. This approach eliminates conflicts at the vertex selection stage by ensuring all user selections remain disjoint, while allowing users complete freedom to perform any transformation operation on their exclusively owned vertex sets (See Fig.~\ref{fig:teaser} (a)).

\paragraph{Action-level Restriction} This preventive approach permits overlapping vertex selections that create NDS conditions but restricts concurrent operations on the joint vertices. When users' selected groups intersect, the system dynamically assigns transformation capabilities based on operation precedence. For instance, if User A initiates rotation on the joint vertices, User B is restricted to different transformation types (translation or scaling) on the overlapping vertices (See Fig.~\ref{fig:teaser} (b)). The system continuously monitors active operations and updates permission matrices in real-time, broadcasting these constraints to all connected clients to maintain consistency.

\paragraph{Reactive Strategies} In this configuration, the system allows both overlapping vertex selections (NDS conditions) and concurrent identical operations on joint vertices, resolving conflicts through computational methods. When multiple users perform the same transformation on joint vertices, the system applies one of three computation schemes to resolve conflicts on the joint vertices while preserving individual transformations on disjoint vertices (e.g.,  Fig.~\ref{fig:teaser} (c)): (i) Additive Combination, (ii) Averaging, or (iii) Intersection. The host processes these calculations based on vertex positions, user grip points, and selected operations, then propagates the resolved transformations to all clients, ensuring synchronized and consistent multi-user manipulation across the shared virtual environment.

\begin{figure}[htb]
  \centering
  \includegraphics[width=1\linewidth]{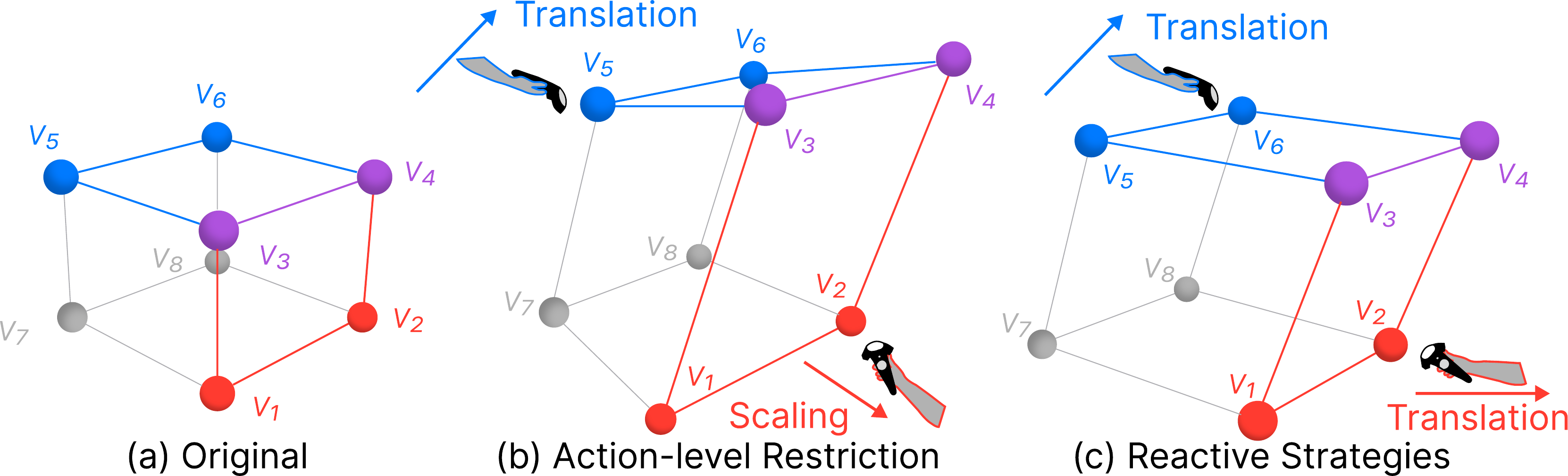}
  \caption{Partial Overlap Support Examples for NDS co-manipulation scenarios.}
  \label{fig:partial}
\end{figure}

While Fig.~\ref{fig:teaser} illustrates scenarios with complete vertex group overlap (joint vertices) for clarity, our system also supports partial overlap (joint and disjoint vertices) between user selections in both Action-level Restriction and Reactive Strategies. When users select non-disjoint vertex groups, the system identifies the joint vertices and computes transformations for these joint vertices relative to their respective selection centroids. Fig.~\ref{fig:partial} shows two examples: (b) two users performing translation and scaling on groups with overlapping vertices separately, and (c) two users performing translation simultaneously. Tab.~\ref{tab:Framework_example} lists the corresponding example operations and outcomes for each conflict resolution strategy in our framework. Cross-referencing Fig.~\ref{fig:partial} with Tab.~\ref{tab:Framework_example} can facilitate a more precise understanding.

\subsection{Operating in Immersive VR without DOF Constraints}

In our system, users can freely perform any transformation operation using the intuitive manipulation methods described in Sec.~\ref{sec:Co-manipulation Function}, leveraging the power of natural interaction with VR controllers and multi-DOF interactions. When conflicts arise from concurrent operations, the system decomposes each user's input into constituent components—translation (x, y, z displacements), rotation (roll, pitch, yaw), and uniform scaling relative to the group centroid. Each DOF component is processed independently according to the strategy before being combined into a unified transformation, thus preserving intuitive VR interactions while effectively managing conflicts.

\subsection{Incorporating Awareness-enhancing Visual Cues}

To enhance collaborative awareness and reduce unintentional conflicts, our system implements a real-time color-based visualization scheme. Unselected vertex groups appear in light blue to indicate availability. Users see their own selections in dark blue and their partner's selections in red, providing a clear distinction of ownership. When selections overlap, the shared vertices are in purple. This visual feedback updates dynamically as users create or modify selections.

\subsection{System Implementation}

The system was developed using Unity (ver 2022.3.27f1) on a computer equipped with a 12th Gen Intel Core i9-14900K processor, 64GB RAM, and an NVIDIA GeForce RTX 4090 GPU, running Windows 11. The participants used the Meta Quest 3 HMD, which supports a resolution of $2064 \times 2208$ pixels and a 120Hz refresh rate. User interactions were implemented using the Meta Interaction SDK, while multi-user networking was facilitated by Unity’s Netcode. Users could join the same virtual environment via the Meta Quest 3 HMD.

\subsection{Study and Task Design}

\begin{figure}[htb]
  \centering
  \includegraphics[width=1\linewidth]{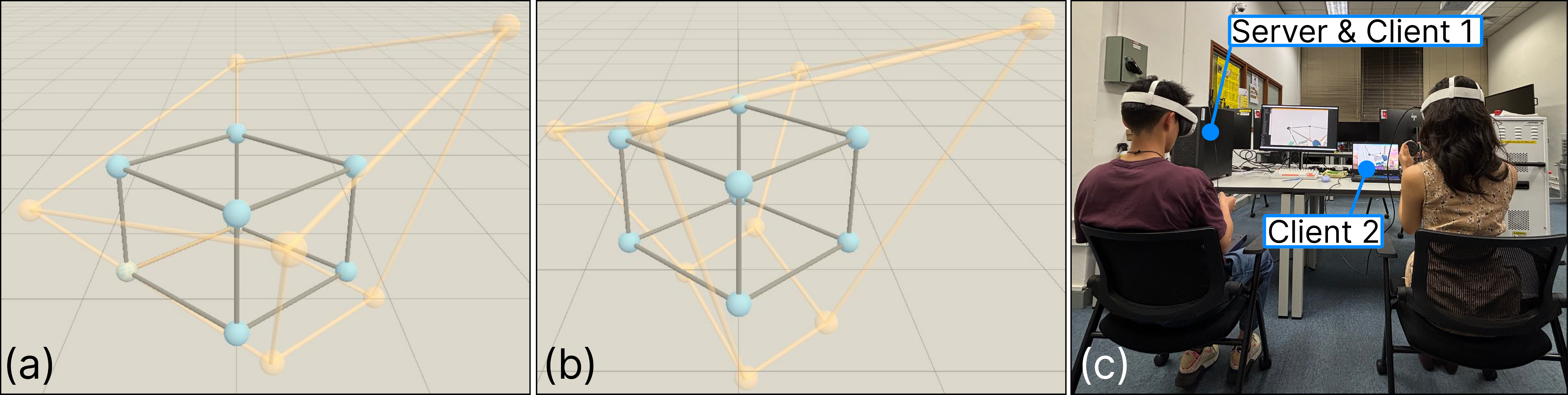}
  \caption{Task environment examples: (a) User Study 1, (b) User Study 2, and (c) the workbench in the physical environment. }
  \label{fig:VE}
\end{figure}

We conducted two user studies to systematically evaluate our conflict resolution framework. \textit{Study 1} examined the \textbf{three computational methods} for Reactive Strategies to determine the optimal approach based on user preference and performance metrics. Based on these findings, \textit{Study 2} compares the three \textbf{conflict resolution strategies}, where Reactive Strategies select the optimal computational method identified in Study 1. Both studies used a standardized model-matching task designed to elicit collaborative manipulation scenarios involving NDS conditions (see Fig.~\ref{fig:VE} (a) \& (b) for examples). We chose this task design because matching wireframe models requires users to manipulate adjacent faces that share common vertices, naturally creating NDS conditions where collaborative conflict resolution becomes necessary. The shared vertices between faces ensure that users cannot simply divide the work independently, forcing them to experience our framework.

Participants were presented with paired wireframe models: (1) an initial configuration consisting of a unit cube ($1m \times 1m \times 1m$) where each vertex could be selected and manipulated, and (2) a semi-transparent target configuration (rendered in light yellow for visual distinction). Target configurations were specifically designed to require the manipulation of adjacent faces that share common vertices. This design creates NDS conditions where users need to collaboratively work on joint vertices while maintaining independent control over disjoint vertices. These target models were pre-generated through the application of transformation (translation, rotation, and scaling) and stored in a library to ensure consistent task complexity across trials. Participants collaboratively manipulated the vertex sets of the initial model using the three transformation operations until achieving spatial alignment with the target. Task completion required satisfying a 5 cm accuracy threshold for all vertices, validated through automatic Euclidean distance computation upon pressing the controller's ``X'' button. The system provided immediate feedback: successful alignment advanced participants to subsequent trials, while deviations exceeding the threshold, a ``Not match'' notification prompting continued refinement.
\section{User Study 1: Method Screening for Optimal Calculation}
\label{sec:Study_1}
To address \textbf{RQ2}, we conducted a within-subjects study comparing three computational methods for Reactive Strategies (Additive Combination, Averaging, and Intersection). Participants completed collaborative model-matching tasks under each computational condition to determine which method optimally balances task efficiency and user experience.

\subsection{Procedure, Measurement, and Participants}
\label{sec:study1_Procedure}
Participants arrived in pre-assigned pairs, completed a demographic questionnaire, and received instruction on the three transformation operations, collaborative matching task, and computational methods. After the researcher helped the participants wear the VR headsets, they underwent training in a simplified test scene, practicing each computational method with single transformation operations. After passing the test scene, the participants removed the headsets for a 1-2 minute break before the formal task.

The formal task required participants to collaboratively transform initial cube models (8 vertices, 12 edges) to match target configurations. We prepared a library of 15 pre-validated target models, each involving transformations of at least two faces with a minimum of two operations per face, ensuring sufficient complexity for meaningful collaboration. For each computational method, 5 models were randomly selected from this library, and participants completed all 5 matching tasks before proceeding to the next method. The presentation order of computational methods was counterbalanced using a Latin square design to alleviate the carryover effects. To ensure conflict situations arose for evaluating the computational methods, participants were instructed to select overlapping vertex groups during manipulation. Performance incentive was provided by ranking all participant pairs, with the top three teams receiving additional compensation. Between computational methods, participants took 10-minute breaks and completed questionnaires assessing their experience. Following the completion of all three conditions, the researcher interviewed participants to discuss their preferences and collaborative experiences. The entire session lasted approximately 90 minutes, and participants received HK\$50 drink vouchers as compensation.

To evaluate user task performance and experience across the three conditions, we employed both quantitative and qualitative analyses. We define concurrent control as instances where both users' selected vertex groups overlap (i.e., NDS co-manipulation). For each pair of participants and each condition, we logged the following objective data for each of the $n_m=5~models$: the completion time for model $i~(t_i)$; the total duration of all concurrent control episodes for model $i~(t_{co,i})$; the total duration within these concurrent episodes during which both participants applied the same operation $(t_{same,i})$; and the number of concurrent control episodes for model $i~(N_{i})$. We also recorded the overall task completion time across all five models $(T)$. Based on these logs, we computed four summary metrics: \textbf{Mean Completion Time} = $\frac{1}{n_m}\sum_{i=1}^{n_m} t_{i}$, which captures the average time to complete a single model; \textbf{Concurrent Time Ratio} = $\frac{1}{T}\sum_{i=1}^{n_m} t_{co,i}$, which is the proportion of the overall task time spent in concurrent control; \textbf{Same-Action Concurrent Ratio} = $\frac{\sum_{i=1}^{n_m} t_{same,i}}{\sum_{i=1}^{n_m} t_{co,i}}$, which is the proportion of concurrent control time during which both participants performed the same operation; and \textbf{Mean Concurrent Duration} = $\frac{\sum_{i=1}^{n_m} t_{co,i}}{\sum_{i=1}^{n_m} N}$, which is the average duration of a single concurrent control episode.

For subjective assessments, we administered the NASA Task Load Index Scale (NASA-TLX) to measure subjective perceived workload~\cite{hart2006nasa}, the User Experience Questionnaire (UEQ) to evaluate user experience with the interactive product~\cite{schrepp2017design}, the System Usability Scale (SUS) to assess system usability~\cite{brooke1996sus}, and the Networked Minds Social Presence Measure to understand the participants' perceptions of partner interaction during the task~\cite{harms2004internal}. At the end of the study, participants were interviewed about their preferences and the reasons behind them.

We recruited 36 participants (18 pairs; 28 females, 8 males) aged 19 -- 31 years (M = 24.33, SD = 2.15) from social media. VR familiarity, measured on a 5-point scale (1 = no familiarity, 5 = high familiarity), averaged 2.42 (SD = 1.11), with 80.56\% having prior VR experience and 13.89\% using VR weekly. Pair familiarity was also measured on a 5-point scale (1 = stranger, 5 = closest friend), averaged 3.31 (SD = 1.80).

\subsection{Results and Discussion}
\label{sec:study1_results}
To evaluate these three computational methods, we conducted one-way repeated measures analysis of variance (RM-ANOVA). Prior to performing RM-ANOVA, we verified the normality of residuals using Shapiro-Wilk tests and examined the sphericity assumption with Mauchly's test. When the sphericity assumption was violated, we applied Greenhouse-Geisser corrections to adjust the degrees of freedom. In cases where the normality assumption was not satisfied, we employed the nonparametric Friedman test as an alternative approach to appropriate data analysis. For post hoc pairwise comparisons, we used paired t-tests when all assumptions were met, and resorted to Wilcoxon signed-rank tests with appropriate multiple comparisons correction when assumptions were violated. These procedures were consistently applied throughout all analyses to ensure the validity and reliability of our findings.

\begin{figure*}[htb]
  \centering
  \includegraphics[width=1\linewidth]{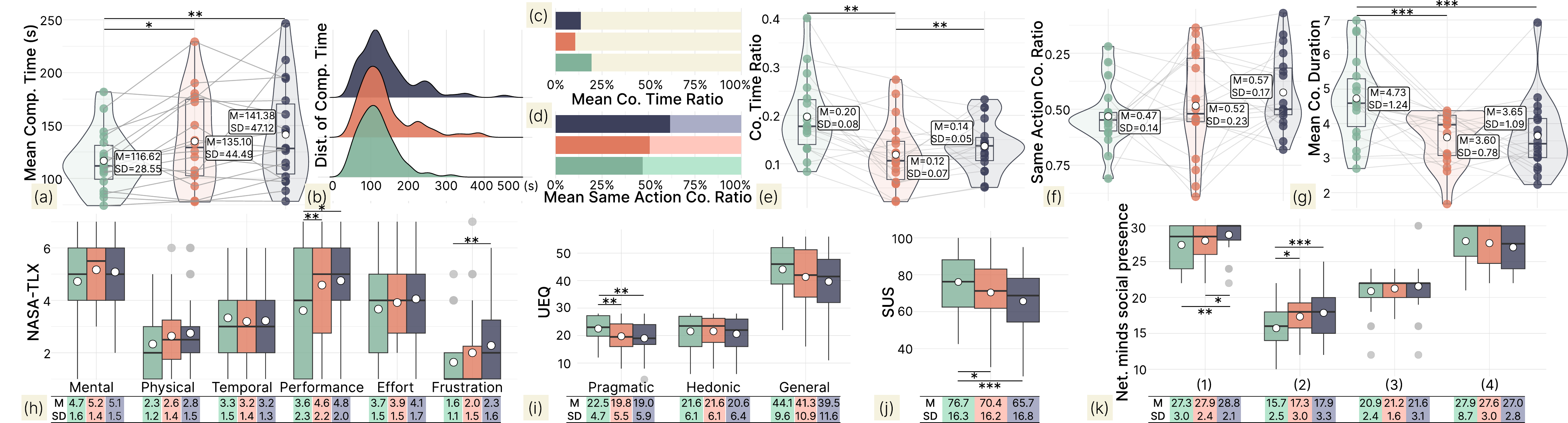}
    \caption{Comparison of Objective Metrics Across Three Methods (a -- g); Box plots of subjective questionnaires for User Study 1 (h -- k). The four subscales of (k) are (1) Co-presence; (2) Attentional allocation; (3) Perceived message understanding, and (4) Perceived behavioral interdependence. The colors represent the \textcolor[HTML]{81b29a}{\rule{1.5ex}{1.5ex}}  \textit{Averaging}, \textcolor[HTML]{e07a5f}{\rule{1.5ex}{1.5ex}} \textit{Additive Combination}, and \textcolor[HTML]{3d405b}{\rule{1.5ex}{1.5ex}} \textit{Intersection} conditon ($p<.05$(*), $p<.01$(**), $p<.001$(***)).}
  \label{fig:Study_1}
\end{figure*}

\paragraph{Task Completion Time Analysis}
\label{sec:Time1}
Task efficiency was assessed by measuring the Mean Completion Time for each pair of participants. Computational methods have a significant effect on completion time ($F_{2,34}=6.00$, $p=.006$, $\eta^2=.065$; Fig.~\ref{fig:Study_1} (a)). Mauchly's test confirmed that the assumption of sphericity was met ($W=.84$, $p=.037$), and both Greenhouse-Geisser and Huynh-Feldt corrections upheld the result. Holm-corrected post-hoc comparisons showed that the \textit{Averaging} method resulted in significantly faster completion times than the \textit{Additive Combination} ($p=.037$) and \textit{Intersection} ($p=.003$) methods. However, no significant differences were observed in the Mean Completion Time between the combine and intersection methods ($p=.483$). To further illustrate differences in task efficiency among methods, Fig.~\ref{fig:Study_1} (b) displays the distribution of completion times per individual model. \textit{Averaging} exhibited a concentrated, left-skewed distribution with minimal variance, while \textit{Additive Combination} and \textit{Intersection} showed broader, right-skewed distributions with greater variability. The \textit{Intersection} method particularly displayed extreme outliers, indicating inconsistent performance across trials. These findings suggest that the \textit{Averaging} method enhances task performance efficiency. The \textit{Averaging} method's superior task execution efficiency may be attributed to its inherent safety when participants inadvertently trigger concurrent manipulations on overlapping vertices. When users unintentionally select overlapping vertices and perform the same operation simultaneously, averaging provides a moderate, predictable outcome that rarely requires correction. Conversely, \textit{Additive Combination} can result in unexpectedly large transformations when concurrent operations compound, potentially disrupting users' intended manipulations. The \textit{Intersection} method's lower efficiency might stem from the difficulty users face in predicting final positions, possibly necessitating frequent adjustments to achieve desired results. These interpretations align with patterns that emerge in the subsequent collaboration analysis and participants' subjective feedback.

\paragraph{Collaboration and Concurrency Analysis}
\label{sec:Action1}
In terms of concurrency control metrics, RM-ANOVA revealed significant effects of method on both Concurrent Time Ratio ($F_{2,34}=10.78$, $p<.001$, $\eta^2=.204$) and Mean Concurrent Duration ($F_{2,34}=10.98$, $p<.001$, $\eta^2=.206$). Sphericity was not violated in either case. Holm-corrected post-hoc tests indicated that the \textit{Averaging} method led to significantly higher Concurrent Time Ratio and longer Mean Concurrent Duration compared to both \textit{Additive Combination} ($p=.005$ and $p=.002$) and \textit{Intersection} ($p=.005$ and $p<.001$), with no significant differences between \textit{Additive Combination} and \textit{Intersection} in either metric. For Same-Action Concurrent Ratio, no significant effect was observed ($F_{2,34}=1.50$, $p=.238$), and no pairwise differences reached significance after Holm correction (all $p>.28$). The \textit{Averaging} method appears to encourage sustained collaborative engagement (this is supported by the results of the Networked Minds Social Presence questionnaire), with users spending more time working concurrently on shared vertices. This increased collaboration likely reflects users' confidence in the predictable, moderated outcomes of averaged transformations. In contrast, the lower concurrent activity in \textit{Additive Combination} and \textit{Intersection} methods suggests that users may have adopted turn-taking strategies to avoid unpredictable results from compound or minimal transformations. Further examination of the data reveals behavioral differences between methods. \textit{Additive Combination} showed polarized patterns. Some pairs immediately switched to sequential control after unexpected large transformations, while others persisted with corrective adjustments. \textit{Intersection} demonstrated more consistent behavior, possibly because its conservative calculation approach, while sometimes producing minimal outputs, remained sufficiently predictable for users to maintain collaborative efforts rather than abandoning concurrent operations entirely. The longer concurrent durations in \textit{Averaging} support this interpretation, suggesting continuous manipulation rather than the brief corrective adjustments characteristic of the other methods.


\paragraph{Subjective Perceived Workload (NASA-TLX)} Significant differences were observed among methods for Mental Demand ($F_{2,70}=3.15$, $p=.049$, $\eta^2=.016$), Physical Demand ($F_{2,70}=3.45$, $p=.037$, $\eta^2=.017$), Performance ($F_{2,70}=6.03$, $p=.004$, $\eta^2=.053$), and Frustration ($F_{2,70}=6.70$, $p=.002$, $\eta^2=.033$). No significant differences were found for Temporal Demand ($F_{2,70}=0.51$, $p=.606$, $\eta^2=.002$) and Effort ($F_{2,70}=2.13$, $p=.126$, $\eta^2=.011$). The Mauchly test indicated that the sphericity was met for all measures (all $p$ values $>.05$). Where minor deviations occurred, Greenhouse-Geisser and Huynh-Feldt corrections were applied, with no changes in significance. Post-hoc pairwise comparisons with Holm corrections revealed significant differences in Performance between \textit{Averaging} and \textit{Additive Combination} ($p=.010$) and between \textit{Averaging} and \textit{Intersection} ($p=.011$). For Frustration, a significant difference was found between \textit{Averaging} and \textit{Intersection} ($p=.010$). No significant pairwise differences were detected for Mental Demand and Physical Demand after Holm correction (all $p$ values $>.05$). These findings highlight that computational methods significantly influenced perceptions of Performance and Frustration, but had limited influence on Mental and Physical Demand, and had no appreciable effect on Temporal Demand and Effort.

\paragraph{User Experience Questionnaire (UEQ)} RM-ANOVA for the UEQ revealed significant effects of the computational method for Pragmatic quality ($F_{2,70}=9.75$, $p<.001$, $\eta^2=.075$) and General user experience ($F_{1.63,57.00}=5.43$, $p=.011$, $\eta^2=.029$), but not for Hedonic quality ($F_{1.49,52.00}=1.13$, $p=.316$, $\eta^2=.005$). Sphericity assumptions were partially violated for Hedonic and General experience; Greenhouse-Geisser corrections were applied accordingly, without changing significance outcomes. Post-hoc pairwise comparisons with Holm corrections indicated significant differences in Pragmatic quality between \textit{Averaging} and \textit{Additive Combination} methods ($p=.006$), and between \textit{Averaging} and \textit{Intersection} methods ($p=.002$). For General user experience, a significant difference was found only between \textit{Averaging} and \textit{Intersection} methods ($p=.013$). No significant pairwise differences were observed for Hedonic quality (all $p$ values $>.05$).

\paragraph{System Usability Scale (SUS)} The analysis revealed a significant effect of computational methods on the SUS scores ($F_{2,70}=10.05$, $p<.001$, $\eta^2=.065$). Post-hoc pairwise comparisons showed significant differences between the \textit{Averaging} method ($M=76.11$, $SD=16.31$) and the \textit{Additive Combination} method ($M=70.42$, $SD=16.20$; $p=.034$), as well as between the \textit{Averaging} and \textit{Intersection} methods ($M=65.69$, $SD=16.75$; $p<.001$). No significant difference was observed between the \textit{Additive Combination} and \textit{Intersection} methods ($p=.067$). Based on industry standards for interpreting SUS scores~\cite{jordan1996usability}, the \textit{Averaging} method (76.11) falls into the good usability range (above 68), indicating that users generally found this method satisfactory and user-friendly. The \textit{Additive Combination} method (70.42) also surpasses the acceptable usability threshold, although it could benefit from further usability improvements. However, the \textit{Intersection} method (65.69) falls slightly below the benchmark (68), indicating potential usability issues requiring immediate attention. The usability ratings align with the objective performance findings. The \textit{Averaging} method's high usability score likely reflects its ability to allow users to maintain a sense of control during concurrent operations. The \textit{Intersection} method's below-threshold score suggests that unpredictable outputs not only hampered efficiency but also degraded the overall user experience.

\paragraph{Networked Minds Social Presence}
\label{sec:presence1}
The Networked Minds Social Presence Questionnaire results further support the evaluation of the methods. Significant effects were found for Co-Presence ($F_{2,70}=8.82$, $p<.001$, $\eta^2=.052$) and Attention Allocation ($F_{2,70}=9.04$, $p<.001$, $\eta^2=.09$). Mauchly’s test indicated a violation of sphericity for Co-Presence ($W=.82$, $p=.038$), but corrected tests confirmed significance. Holm-corrected post-hoc tests showed that for Co-Presence, \textit{Intersection} was rated significantly higher than \textit{Averaging} ($p=.004$) and \textit{Additive Combination} ($p=.014$), with no significant difference between \textit{Averaging} and \textit{Additive Combination} ($p=.070$). For Attention Allocation, both \textit{Additive Combination} ($p=.013$) and \textit{Intersection} ($p<.001$) were rated significantly higher than \textit{Averaging}. No significant effects were observed for Perceived Message Understanding ($F_{2,70}=.97$, $p=.386$) or Perceived Behavioral Interdependence ($F_{2,70}=1.75$, $p=.182$), indicating that these dimensions were not affected by the computational method. The \textit{Intersection} method's higher Co-Presence ratings paradoxically suggest that unpredictability may enhance awareness of one's collaborator, albeit potentially through interference rather than seamless cooperation. The elevated Attention Allocation scores for both \textit{Intersection} and \textit{Additive Combination} indicate that participants needed to monitor their partner's actions more closely, possibly to anticipate or compensate for unpredictable outcomes. Conversely, \textit{Averaging}'s lowest Attention Allocation score suggests it enabled participants to maintain task focus while remaining peripherally aware of their partner, facilitating what might be considered more natural, less cognitively demanding collaboration.

\paragraph{Preferences and Subjective Comments} The vast majority of the participants (N = 31, 86.11\%) rated \textit{Averaging} as the best method, whereas more than half of the participants (N = 23, 63.89\%) thought \textit{Intersection} was the worst method. In addition, some participants (N = 5, 13.89\%) considered that \textit{Additive Combination} was the best condition. However, none of the participants thought that \textit{Intersection} was the best method. We scored the three calculation method according to the preferences of the participants (1: Worst - 3: Best), the results of Friedman test indicated significant differences in preferences, $\chi^2$(2) = 43.167, $p < .001$, with a large effect size (Kendall's $W = .600$). Post-hoc analysis using the Nemenyi test revealed that \textit{Averaging} (M rank = 2.861) was significantly preferred over both \textit{Additive Combination} (M rank = 1.778) and \textit{Intersection} (M rank = 1.361), while no significant difference between \textit{Additive Combination} and \textit{Intersection}.

We asked the participants for their reasons for ranking the three calculations and, interestingly, participants who favored the \textit{Additive Combination} method consistently cited its conceptual simplicity and alignment with physical intuition, particularly the force addition analogy, noting that coordinated actions in the same direction could accelerate task completion. However, others found that direct summation led to unexpectedly large transformations and loss of control, especially when users inadvertently performed simultaneous operations or struggled to anticipate their partner's actions due to differing viewpoints. In contrast, the \textit{Averaging} method emerged as the most consistently acceptable approach, with no participants ranking it last. While some described it as ``\textit{slow}'', ``\textit{gentle}'', and ``\textit{conservative}'', these characteristics were often viewed positively when precise manipulation is required, with participants finding it ``\textit{safer}'' for precise manipulation tasks. The \textit{Intersection} method received predominantly negative feedback, with participants characterizing the results as ``\textit{jumpy}'', ``\textit{unpredictable}'', and ``\textit{out of control}'', fundamentally disrupting collaborative dynamics to the extent that one participant remarked it ``\textit{make me feel we are not teammates but opponents}''. These subjective assessments corroborate our quantitative findings, revealing that predictability and control stability are important in collaborative manipulation. The \textit{Averaging} method appears to best support effective co-manipulation, while \textit{Intersection} creates adversarial rather than collaborative interaction, although it increases awareness of partners.

\subsection{Key Implications}
\label{sec:Study_1_Implications}
The evaluation in Study 1 provided critical insights that informed our subsequent design of the conflict resolution strategy. The \textit{Averaging} method demonstrated better performance in multiple dimensions: highest task efficiency, sustained collaborative engagement, good usability (SUS = 76.11), and minimal attention allocation demands, allowing seamless collaboration. These converging results establish \textit{Averaging} as the primary calculation method for reactive strategies. Additionally, while averaging provides stability, participant feedback revealed a notable limitation: its moderation effect can impede rapid corrective actions. To address this, we introduce a ``second-user priority'' approach where concurrent operations are resolved by accepting only the most recent input, effectively allowing immediate corrections without averaging delays. Consequently, Study 2 evaluates four distinct conflict resolution strategies: two preventive approaches (\textit{Object-level Restriction} and \textit{Action-level Restriction}) that avoid conflicts through access control, and two reactive approaches (\textit{Reactive Strategies I} using averaging and \textit{Reactive Strategies II} using second-user priority) that computationally resolve concurrent operations. This design enables a comprehensive comparison between conflict prevention and resolution paradigms in multi-object collaborative manipulation.
\section{User Study 2: Comparative Analysis of Conflict Resolution Solutions}

Based on the findings of Study 1, we address \textbf{RQ3} by comparing different conflict resolution strategies.

\subsection{Participants, Study Design and Measurement}
Another 40 participants (20 pairs; 29 females, 11 males) were recruited from a social media platform, and the participants were aged 19 to 30 years (M = 24.625, SD = 3.19). As none of these participants had taken part in Study 1, no learning or carryover effects between the two studies exist. Their mean value of VR familiarity was 2.30 (SD = 1.09) on a 5-point scale. 65\% of the participants had previously used the VR device. The average familiarity of the pair was 4.00 (SD = 1.15). We followed Study 1's within-subjects design but increased task complexity to more reliably trigger NDS conditions. Each model now contained more intersecting planes that needed to be manipulated with joint vertices, requiring participants to manipulate more overlapping vertex groups and perform more operations. Specifically, each model required transformations across at least three faces (compared to two in Study 1), each involving all three operations. Due to this increased complexity per model, we reduced the number of models from five to three per condition to maintain a reasonable session duration. Participants completed these three models per condition (Object-level Restriction [\textit{OLR}], Action-level Restriction [\textit{ALR}], Reactive Strategies I [\textit{RS\_I}; averaging], Reactive Strategies II [\textit{RS\_II}; second-user priority]). The entire session lasted approximately 120 minutes, and each participant received HK\$100 drink vouchers as compensation. We maintained Study 1's procedures, metrics, and questionnaires (see Sec.~\ref{sec:study1_Procedure}).

\subsection{Results and Discussion}
\label{sec:study2_results}
The second user study follows the same data analysis process as Study 1, with details given at the beginning of Sec.~\ref{sec:study1_results}.

\begin{figure*}[htb]
  \centering
  \includegraphics[width=\linewidth]{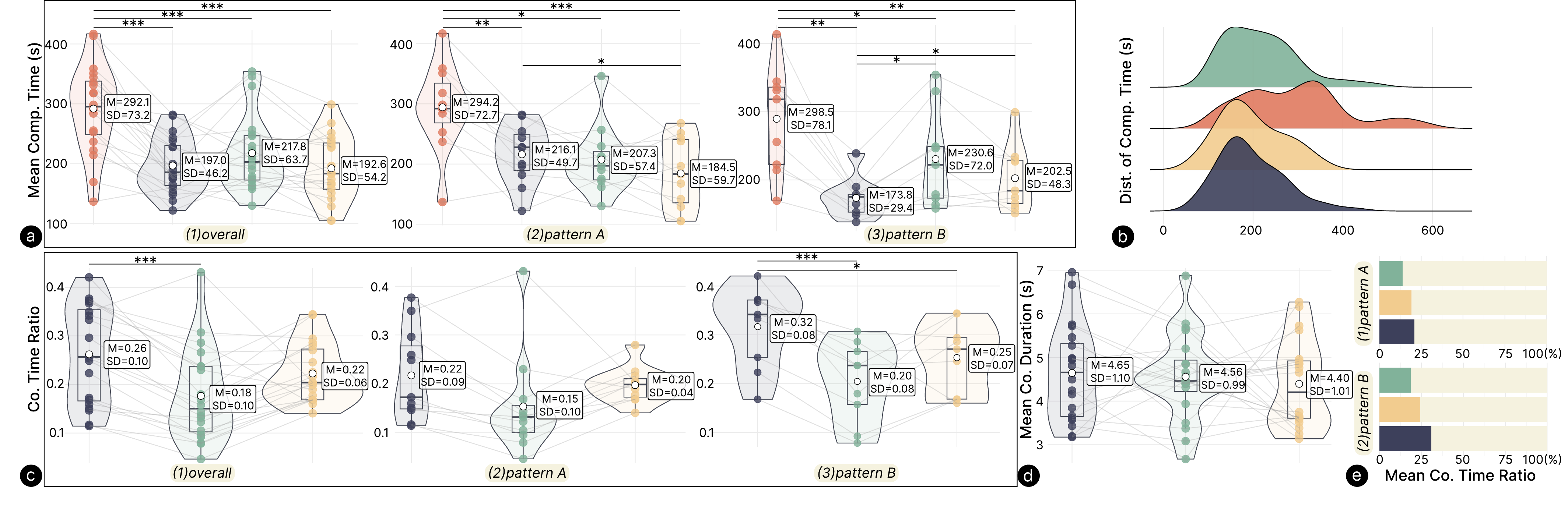}
  \caption{Comparison of Objective Metrics Across Different Conflict Resolution Strategies. The colors represent the \textcolor[HTML]{e07a5f}{\rule{1.5ex}{1.5ex}}  \textit{Object-level Restriction (OLR)}, \textcolor[HTML]{3d405b}{\rule{1.5ex}{1.5ex}} \textit{Action-level Restrictions (ALR)}, \textcolor[HTML]{81b29a}{\rule{1.5ex}{1.5ex}} \textit{Reactive Strategies II (RS\_II)} and \textcolor[HTML]{f2cc8f}{\rule{1.5ex}{1.5ex}} \textit{Reactive Strategies I (RS\_I)} conditions ($p<.05$(*), $p<.01$(**), $p<.001$(***)).}
  \label{fig:Task_completion_time_2}
\end{figure*}

\paragraph{Task Completion Time Analysis} During the study, we observed that participants naturally adopted one of two distinct collaboration strategies. \textbf{Pattern A} followed sequential assistance with \textbf{clear task division}: participants initially divided the workload and worked independently (avoid overlap of groups) in their assigned groups, after which the faster participant provided assistance to help complete the remaining groups of their partner. In contrast, \textbf{Pattern B} employed a collaboration approach \textbf{without predefined task division}: participants freely selected and modified any vertex groups without coordination, regardless of whether their selections overlapped with their partner's. In general, an RM-ANOVA revealed a significant main effect of Method on Mean Completion Time ($F_{3,67}=27.12$, $p<.001$, $\eta^2=.317$). Mauchly's test confirmed that the sphericity assumption was met ($W=0.698$, $p=0.272$). The \textit{OLR} condition ($M=291.85s$) showed significantly longer completion times compared to \textit{ALR} ($p<.001$), \textit{RS\_II} ($p<.001$), and \textit{RS\_I} ($p<.001$) conditions. However, no significant differences emerged between the \textit{ALR}, \textit{RS\_II}, and \textit{RS\_I} conditions (all $p>.10$). However, if we group users' performance according to different patterns of collaboration, we find for the \textbf{Pattern A}, the analysis revealed a significant main effect of Method ($F_{3,30}=16.90$, $p<.001$, $\eta^2=.340$). Post-hoc comparisons showed that \textit{OLR} significantly differed from all other conditions (all $p<.012$), while \textit{ALR} vs. \textit{RS\_I} also reached significance ($p=.033$). A significant main effect also exists for \textbf{Pattern B} ($F_{3,24}=16.90$, $p<.001$, $\eta^2=.362$), Post-hoc analysis revealed that \textit{OLR} differed significantly from \textit{ALR} ($p=.004$) and \textit{RS\_I} ($p=.008$), while \textit{ALR} vs. \textit{RS\_II} and \textit{ALR} vs. \textit{RS\_I} comparisons also reached significance (both $p=.033$). Fig.~\ref{fig:Task_completion_time_2} (b) presents the probability density distributions of completion times across all participants for each condition. The \textit{OLR} condition exhibits the broadest distribution (150-400+ seconds) with high variability and a pronounced right tail, indicating that some participants experienced substantially longer completion times. In contrast, the \textit{ALR} and \textit{RS\_I} conditions show the most concentrated distribution centered around 175 seconds with minimal variance, suggesting consistent performance across users. The \textit{RS\_II} condition displays intermediate distributions with moderate spread, showing improvement compared to \textit{OLR}.

\paragraph{Collaboration and Concurrency Analysis} We examined two collaboration metrics to understand how conflict resolution design influenced collaborative behavior: Concurrent Time Ratio and Mean Concurrent Duration. An RM-ANOVA revealed a significant main effect of the condition on Concurrent Time Ratio ($F_{2,38}=11.45$, $p<.001$, $\eta^2=.144$), with post-hoc pairwise comparisons showing \textit{ALR} significantly differed from \textit{RS\_II} condition ($p<.001$), while \textit{RS\_I} condition showed no significant differences from either \textit{ALR} or \textit{RS\_II} (both $p = .076$). In contrast, Mean Concurrent Duration showed no significant differences across conditions ($F_{2,38}=0.28$, $p=.757$), indicating consistent collaboration episode lengths regardless of conflict design. Based on the analysis of the different collaboration patterns, we found that while \textbf{Pattern A} showed only marginally significant effects for Concurrent Time Ratio ($F_{2,20}=3.22$, $p=.061$), \textbf{Pattern B} demonstrated a more substantial significant effect ($F_{2,16}=10.85$, $p<.001$, $\eta^2=.272$), with \textit{ALR} vs. \textit{RS\_II} comparison reaching high significance ($p<.001$). These findings suggest that the conflict design condition significantly influenced the proportion of time participants spent in active collaboration, with \textit{ALR} consistently promoting higher concurrent ratios compared to \textit{RS\_II} condition, particularly in continuous collaboration workflows (\textbf{Pattern B}). However, since the duration of individual collaboration episodes remained stable across conditions, the results indicate that conflict design affects the frequency or initiation of collaborative interactions rather than their sustained duration.
 
\begin{figure*}[htb]
  \centering
  \includegraphics[width=1\linewidth]{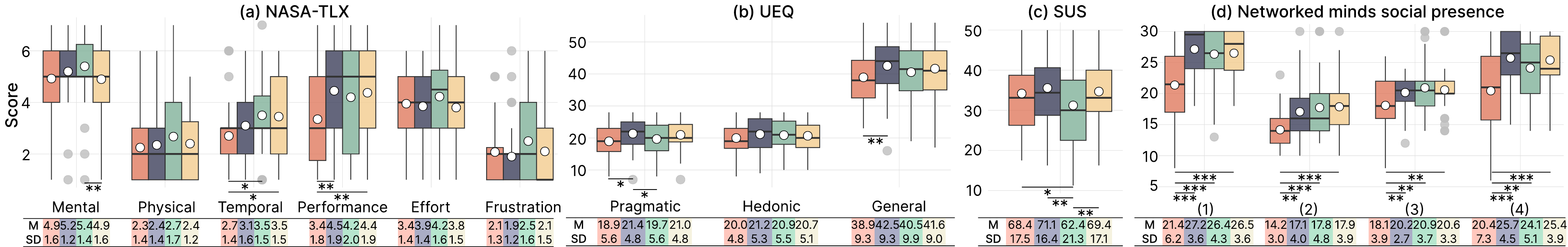}
  \caption{Box plots of subjective questionnaires for User Study 2. The four subscales of (d) are (1) Co-presence; (2) Attentional allocation; (3) Perceived message understanding, and (4) Perceived behavioral interdependence. The colors represent the \textcolor[HTML]{e07a5f}{\rule{1.5ex}{1.5ex}}  \textit{Object-level Restriction (OLR)}, \textcolor[HTML]{3d405b}{\rule{1.5ex}{1.5ex}} \textit{Action-level Restrictions (ALR)}, \textcolor[HTML]{81b29a}{\rule{1.5ex}{1.5ex}} \textit{Reactive Strategies II (RS\_II)} and \textcolor[HTML]{f2cc8f}{\rule{1.5ex}{1.5ex}} \textit{Reactive Strategies I (RS\_I)} conditon ($p<.05$(*), $p<.01$(**), $p<.001$(***)).} 
  \label{fig:Study_2}
\end{figure*}

\paragraph{Subjective Perceived Workload (NASA-TLX)} RM-ANOVA revealed significant main effects of condition on Mental Demand ($F_{3,117}=2.69$, $p=.050$, $\eta^2=.020$), Temporal Demand ($F_{3,117}=4.69$, $p=.003$, $\eta^2=.045$), Performance ($F_{3,117}=6.18$, $p<.001$, $\eta^2=.051$), and Frustration ($F_{3,117}=3.54$, $p=.017$, $\eta^2=.025$). Mauchly’s test indicated violations of sphericity for several subscales; thus, Greenhouse-Geisser and Huynh-Feldt corrections were applied as necessary. The corrected p-values confirmed the significance for all four subscales. No significant effects were found for Physical Demand ($F_{3,117}=1.92$, $p=0.131$) or Effort ($F_{3,117}=1.15$, $p=0.332$). Holm-corrected post-hoc comparisons for Mental Demand revealed a significant difference only between \textit{RS\_II} and \textit{RS\_I} ($p=.010$). For Temporal Demand, significant differences were found between \textit{OLR} and both \textit{RS\_II} ($p=.024$) and \textit{RS\_I} ($p=.024$). For Performance, \textit{OLR} was rated significantly higher than both \textit{ALR} ($p=.010$) and \textit{RS\_I} ($p=.003$). Finally, regarding Frustration, no other pairwise comparisons reached significance after Holm correction.

\paragraph{User Experience Questionnaire (UEQ)} RM-ANOVA on the UEQ revealed a significant main effect of condition on Pragmatic Quality ($F_{3,117}=5.04$, $p=.003$, $\eta^2=.034$) and General User Experience ($F_{3,117}=5.03$, $p=.003$, $\eta^2=.021$). Mauchly’s test indicated violations of sphericity for both scales, and Greenhouse-Geisser and Huynh-Feldt corrections confirmed the significance of these effects. No significant effect was observed for Hedonic Quality ($F_{3,117}=1.72$, $p=.167$). Holm-corrected post-hoc comparisons showed that \textit{ALR} was rated significantly higher than \textit{OLR} in both Pragmatic Quality ($p=.012$) and General UEQ score ($p=.006$), and significantly higher than \textit{RS\_II} in Pragmatic Quality ($p=.044$). No other pairwise comparisons reached significance following correction. These results suggest that the \textit{ALR} condition was perceived as more positively pragmatic than the \textit{OLR} and \textit{RS\_II}, while hedonic impressions remained unaffected across conditions.

\paragraph{System Usability Scale (SUS)} The analysis revealed a significant main effect of condition on the SUS scores ($F_{3,117}=7.10$, $p<.001$, $\eta^2=.032$). Mauchly’s test indicated a violation of the sphericity assumption ($w=.74$, $p=.042$); therefore, Greenhouse-Geisser and Huynh-Feldt corrections were applied, and both confirmed the significance of the effect. Holm-corrected post-hoc comparisons revealed that the \textit{RS\_II} condition was rated significantly lower than both \textit{ALR} ($p=.006$) and \textit{RS\_I} ($p=.006$), and also significantly lower than \textit{OLR} ($p=.035$). No significant differences were observed among the remaining conditions. Descriptive statistics indicated that \textit{ALR} received the highest mean SUS score ($M=71.13$, $SD=16.36$), followed by \textit{RS\_I} ($M=69.38$, $SD=17.06$), \textit{OLR} ($M=68.38$, $SD=17.48$), and \textit{RS\_II} ($M=62.44$, $SD=21.26$). These results suggest that while all conditions achieved moderate usability ratings, the \textit{RS\_II} condition was perceived as notably less usable relative to the others. These results indicate that the mean scores for all conditions were at the average level of the SUS scale except for \textit{RS\_II} condition.

\paragraph{Networked Minds Social Presence} The results showed significant main effects of condition on Co-Presence ($F_{3,117}=21.73$, $p<.001$, $\eta^2=.208$), Attention Allocation ($F_{3,117}=10.18$, $p<.001$, $\eta^2=.127$), Perceived Message Understanding ($F_{3,117}=6.30$, $p<.001$, $\eta^2=.095$), and Perceived Behavioral Interdependence ($F_{3,117}=13.21$, $p<.001$, $\eta^2=.136$). Mauchly’s test identified sphericity violations in the Behavioral Interdependence dimension; Greenhouse-Geisser and Huynh-Feldt corrections were applied where necessary, with all corrected p-values confirming significance. Holm-corrected post hoc tests revealed that the \textit{OLR} condition was rated significantly lower than \textit{ALR}, \textit{RS\_II}, and \textit{RS\_I} for Co-Presence ($p<.001$ for all), Attention Allocation ($p=.001$, $p<.001$, and $p<.001$, respectively), and Perceived Message Understanding ($p=.007$, $p=.007$, and $p=.006$, respectively). Additionally, for Perceived Behavioral Interdependence, \textit{OLR} scored significantly lower than \textit{ALR} ($p<.001$), \textit{RS\_II} ($p=.002$), and \textit{RS\_I} ($p<.001$). No other pairwise comparisons reached significance. These results indicate that all experimental conditions (\textit{ALR}, \textit{RS\_II}, and \textit{RS\_I}) enhanced users’ perceived social presence across key dimensions compared to the \textit{OLR} condition.

\paragraph{Preferences and Subjective Comments} The majority of participants found \textit{ALR} (N = 21, 52.5\%) or \textit{RS\_I} (N = 15, 37.5\%) to be the best experience, the vast majority of participants found \textit{OLR} (N = 31, 77.5\%) to be the worst experience, and no participants considered \textit{ALR} to be the worst strategy. Friedman test ($\chi^2$(3) = 63.630, $p < .001$; Kendall's $W = .530$) and Post-hoc analysis revealed a clear hierarchy: \textit{ALR} (M rank = 3.375) and \textit{RS\_I} (M rank = 3.150) were both significantly preferred over \textit{RS\_II} (M rank = 2.125) and \textit{OLR} (M rank = 1.350), while \textit{ALR} and \textit{RS\_I} did not differ significantly from each other. \textit{RS\_II} was also significantly preferred over \textit{OLR}. 

\textit{ALR} received positive feedback for being ``\textit{not disturbed}'' and ``\textit{less likely to have uncontrollable conflicts}'', with some participants feeling it was suitable for precision tasks. However, some found it restrictive, noting frustration when they ``\textit{cannot move sometimes}'' due to operational conflicts. The \textit{RS\_I} method garnered appreciation for its collaborative nature, with participants valuing the ability to help adjust each other's actions (``\textit{I don't have to worry about moving too much, my friend can help me to adjust together}''), though others criticized that the averaged result might satisfy neither user's intentions. The \textit{RS\_II} strategy proved particularly polarizing: experienced collaborators praised its efficiency for quick corrections (``\textit{I can use it to quickly adjust it back for him}''), while others found it unpredictable and disruptive, with sudden partner interventions causing results to ``fly off the handle somehow''. The \textit{OLR} condition received predominantly negative feedback, characterized as ``\textit{boring}'' and ``\textit{inefficiency}'', with participants frustrated by forced waiting and the cumbersome process of group reassignment. One participant noted, ``\textit{I honestly didn't realize that it was two people playing a game}'', highlighting the method's failure to support meaningful collaboration. However, two participants preferred \textit{OLR}'s complete independence from partner interference. One participant suggested that different strategies served distinct purposes, and both \textit{RS\_I} and \textit{RS\_II} ``\textit{could both exist in the system for different purposes}'', indicating that optimal conflict resolution may require context-dependent strategy selection rather than a one-size-fits-all approach.
\section{Discussion}

\linkedsubsection{RQ}{Extending Conflict Resolution to Multi-object (RQ1)}
Existing research on co-manipulation specializes in whole-object manipulation, treating virtual objects as monolithic entities (see Sec.~\ref{sec:Design_Opportunities}). However, this limitation fails to support VR applications where users manipulate sub-components. By identifying five design opportunities from this gap, we developed a comprehensive framework that extends conflict resolution to NDS conditions. We first propose an intuitive mechanism to manipulate sub-components in immersive VR. By implementing a group-based hierarchy anchored at vertex group centroids, our approach maintains spatial relationships during manipulation. This concept aligns with Xia et al.'s ``container'' metaphor for collaborative editing~\cite{Xia2018Spacetime}. This design was acceptable to participants with limited VR experience (mean familiarity = 2.36/5.0 in both studies), the above-average SUS scores under most conditions (except \textit{RS\_II}) validate the usability of our transformation operations approach. While the \textit{OLR} in our framework represents the current standard for multi-user VR applications~\cite{tiltbrush,gravitysketch}, our evaluation in User Study 2 highlights its limitations (more details are discussed in Sec.~\ref{sec:dis_RQ3}). Future developers can apply \textit{ALR} and \textit{Reactive Strategies} within collaborative or multiplayer applications involving sub-components. For example, current VR painting applications (e.g., Tilt Brush~\cite{tiltbrush}, Quill~\cite{quill}) use whole-canvas locking (similar to \textit{OLR}), preventing simultaneous edits. With our framework, \textit{ALR} would allow artists to work on different brushstroke groups while preventing conflicts when modifying shared strokes. \textit{Reactive Strategies} would enable smooth blending when artists intentionally work on the same area, creating collaborative effects. Particularly during early ideation phases, where artists explore rough concepts and engage in creative ``idea collisions'', the Averaging method can preserve contributions from all collaborators rather than allowing one vision to dominate. In addition, current VR modeling tools (e.g., Gravity Sketch~\cite{gravitysketch}, Medium~\cite{medium}) often force users to take turns when editing the same model. Our framework enables precision work, allowing users to simultaneously edit different parts of a model (e.g., \textit{ALR}) and make quick corrections by experts during iterative design reviews (e.g., \textit{RS\_II}).

Our framework provides a foundation that can be extended in several ways. For instance, the set of permissible actions could extend beyond the three basic transformations we implemented (see Sec.~\ref{sec:limitation}). Similarly, the computational methods in \textit{Reactive Strategies} need not be limited to the specific approaches proposed in our work; developers could implement computations that are more directly relevant to their application domain, even transcending purely algebraic operations (e.g., rendering of materials or lights in virtual environments). 


\linkedsubsection{RQ}{Computational Method for Reactive Strategies (RQ2)}

Our evaluation revealed that the \textit{Averaging} method emerged as the optimal computational approach for reactive conflict resolution in multi-object co-manipulation. This may be because the \textit{Averaging} method's conservative transformation approach provided predictable and stable outcomes that participants could readily anticipate and control. In multi-object scenarios, users must divide their attention across multiple vertices within their selected groups, often focusing primarily on the grabbed vertex that serves as the manipulation handle. This distributed attention makes it challenging to maintain awareness of the entire model's state (see NASA-TLX in Sections~\ref{sec:study1_results} and \ref{sec:study2_results}, where all conditions require high mental demand). The \textit{Averaging} method's moderated transformations compensated for this limited awareness by preventing drastic unexpected changes, thereby reducing cognitive load and maintaining user confidence. In contrast, the \textit{Additive Combination} method, while conceptually aligned with physical force addition, revealed problems in our user study. The direct summation of concurrent transformations often resulted in excessive displacements that exceeded users' intended movements, particularly when participants inadvertently triggered simultaneous operations. While some experienced users with strong spatial abilities found it efficient and physically intuitive, the majority found the directly additive transformations disruptive to precise manipulation tasks. The \textit{Intersection} method performed poorest, contradicting findings from previous AR and 2D screen-based studies~\cite{Grandi2018Design,Ruddle2002Symmetric}. This may be due to the difference in the interaction paradigms. Prior work typically constrained manipulations to a single DOF, simplifying the mental calculation of intersection outcomes. However, in immersive VR with 6-DOF controllers, users must simultaneously consider intersections across multiple transformation dimensions (3-DOF for translation, 3-DOF for rotation). This computational complexity, compounded by the multi-object context where intersections must be calculated across numerous vertices, resulted in unpredictable outputs that participants characterized as causing loss of control.

\linkedsubsection{RQ}{Comparing Preventive vs. Reactive Strategies (RQ3)}
\label{sec:dis_RQ3}
Our comparative analyses found trade-offs between preventive and reactive conflict resolution approaches in NDS co-manipulation. The \textit{OLR} strategy, despite preventing conflicts through exclusive object locking, demonstrated poor performance on both objective and subjective measures. Its significantly longer task completion times and lowest social presence scores indicate that preventing conflicts through rigid access control destroys collaboration. This finding is consistent with the results of the social presence questionnaire, which suggests that exclusive locking tends to transform collaborative tasks into isolated work, thereby avoiding conflict and eliminating the benefits of co-manipulation~\cite{Greenberg1994Real}. In contrast, \textit{ALR} emerged as the most successful strategy overall, achieving the highest usability scores and pragmatic quality ratings while maintaining task efficiency. This suggests that restricting the types of operation that users can perform simultaneously, rather than the objects they can select, can maintain collaborative participation while preventing destructive conflict. The strategy's highest Concurrent Time Ratio indicates that users actively collaborate under this condition. In addition, participants commented ``not disturbed'' and ``less likely to have uncontrollable conflicts'', suggesting it may be suitable for precision collaborative tasks. This is consistent with the findings of Wieland et al.~\cite{Wieland2021Separation}. However, \textit{ALR} was not without limitations; some participants expressed frustration with operational restrictions, preferring unrestricted access to all transformation types when needed. This potentially reflects the different preferences for strategies between users who prioritize precise operations (prefer \textit{ALR}) and those seeking a smoother collaborative experience (prefer \textit{RS\_I}). Future applications need to choose the most appropriate conflict resolution strategy for VR collaboration.

The influence of collaboration patterns provides additional insights. Users with \textbf{Pattern A} performed better under reactive strategies, particularly \textit{RS\_I}, while users with \textbf{Pattern B} achieved better results under \textit{ALR}. 
This may be attributed to the fact that users in \textbf{Pattern A} who complete their assigned tasks first tend to be more skilled and spatially aware, thus preferring the flexibility of reactive strategies to assist their partners without restrictions. 
Conversely, users with \textbf{Pattern B} that lacks prior task division can benefit from \textit{ALR}'s tighter constraints to prevent unexpected concurrent behaviors. 
This interpretation is supported by a higher Concurrent Time Ratio compared to \textit{RS\_I} and \textit{RS\_II} observed in \textbf{Pattern B}. Noticeably, \textbf{Pattern A} showed no significant differences across conditions.
These findings suggest that collaboration styles could potentially influence the way we recognize the optimality of conflict resolution strategies. Therefore, further research is needed to investigate these interaction effects in greater detail. 

\subsection{Design Implications}

Our study generates design guidance, and we summarize four design implications for conflict resolution strategies in co-manipulation within VR collaboration. 

\textit{\textbf{DI1:} Provide Strategy Selection Based on Task Requirements.} Users should be able to switch between strategies for trade-offs between precision control (e.g., \textit{ALR} that is advantageous for CAD modeling) and seamless collaboration (e.g., \textit{Reactive Strategies} with averaging for conceptual design or brainstorming). That is, the task nature drives the needs of co-manipulation accuracy or collaborative workflow preferences.

\textit{\textbf{DI2:} Adapt to User Collaboration Patterns.}  As we observed users following \textbf{Pattern A} in User Study 2, \textit{Reactive Strategies} offer flexibility for helping partners complete tasks, while users preferring \textbf{Pattern B} benefit from \textit{ALR}'s inherent concurrency constraints. Systems should detect and suggest appropriate strategies based on observed collaboration styles.

\textit{\textbf{DI3:} Consider User Experience and Skill Level.} Novice users should default to \textit{ALR} (highest SUS scores, lowest frustration) to build confidence, while experienced teams can leverage \textit{RS\_II} for rapid corrections despite its higher mental demands. The system design should consider the skill and experience levels of different users.

\textit{\textbf{DI4:} Balance Awareness and Control Trade-offs.} While \textit{OLR} prevents conflicts entirely, it significantly impairs collaboration (lowest social presence scores, longest completion times). Systems should favor strategies that maintain collaborative awareness (\textit{ALR} or \textit{RS\_I}) over complete isolation of users. Visual cues or haptic feedback can potentially compensate for reduced control predictability in reactive approaches.

\subsection{Limitations and Future Work}
\label{sec:limitation}
Our study has several limitations, and accordingly, we suggest directions for future research. First, participants recruited from local social media had limited VR experience, which may have influenced strategy preferences. For example, experienced users might prefer \textit{Reactive Strategies} for their flexibility, while novices may favor \textit{ALR} for its simplicity. Future work should examine how VR expertise affects collaborative behavior and strategy effectiveness. Wieland et al.~\cite{Wieland2021Separation} focused exclusively on pairs who were previously acquainted, our work addresses this limitation by including working collaborators who were unrecognized (strangers). However, we did not explicitly analyze the level of familiarity between pairs in detail. Future work should explore this aspect, as varying levels of familiarity may influence collaboration patterns and consequently affect the effectiveness of different conflict resolution strategies~\cite{yu2021familiarity,janssen2009influence}. Furthermore, in User Study 2, we observed two distinct collaboration patterns (Pattern A: with task division, Pattern B: without task division) that emerged naturally during the study. These patterns significantly influenced strategy effectiveness, with Pattern A users performing better with \textit{Reactive Strategies} and Pattern B users achieving better results with \textit{ALR}. However, these represent only a subset of possible collaboration styles that may exist in collaborative VR environments. Therefore, future research should investigate how different collaboration approaches and task allocation affect the preferences and performance of the conflict resolution strategy.

From a technical perspective, we evaluated only wireframe model editing, focusing on basic geometric transformations. While this setting provided controlled testing conditions, real-world applications involve more complex geometries and diverse manipulation requirements. As such, future studies should validate our framework across different domains, including emerging techniques such as \textit{3D Gaussian Splatting} where each Gaussian exists as an individually manipulable element similar to our wireframe vertices~\cite{GaussianSplatting}, with advanced transformation operations beyond translation, rotation, and scaling (e.g., boolean operations, extrusion of faces or edges, surface smoothing). Additionally, our evaluation was limited to two-user scenarios. In contrast, multi-user collaboration with three or more participants introduces additional challenges in conflict detection, resolution priority, and user awareness that require investigation. Indeed, the scalability of our strategies with increased user count remains unexplored. Finally, simple second-user priority scheme resolves conflicts efficiently in our two-user setting, but may disrupt other collaboration patterns or larger teams; exploring richer schemes (e.g., role-based or negotiated priorities) is a promising direction.
\section{Conclusion}
In this article, we presented a conflict resolution framework for the non-disjoint set (NDS) condition in immersive VR environments. Through evaluation in two user studies, we identified optimal computational approaches and compared preventive versus reactive conflict resolution strategies. Our first study indicated that the Averaging method provides the best balance between task efficiency and user experience among reactive computational approaches, offering predictable results while maintaining collaborative engagement. Building on these findings, our second study demonstrated that Action-level Restriction achieves better usability and a more collaborative experience compared to Object-level Restriction, while Reactive Strategies, especially Averaging, support smoother collaboration for experienced users, second-user priority is better suited for quick corrections. We observed that optimal strategy selection depends on multiple factors, including task requirements, user experience, and collaboration patterns. Our framework and design implications provide guidance for developers designing collaborative VR editing systems.






\bibliographystyle{IEEEtran}
\bibliography{main.bib}

@article{8ca911d754f448fdb959f34677db6f95,
title = "All One Needs to Know about Metaverse: A Complete Survey on Technological Singularity, Virtual Ecosystem, and Research Agenda",
author = "Lee, {Lik Hang} and Tristan Braud and Zhou, {Peng Yuan} and Lin Wang and Dianlei Xu and Zijun Lin and Abhishek Kumar and Carlos Bermejo and Pan Hui",
note = "Publisher Copyright: {\textcopyright}2024 L.-H. Lee et al.",
year = "2024",
month = nov,
day = "6",
doi = "10.1561/1100000095",
language = "English",
volume = "18",
pages = "100--337",
journal = "Foundations and Trends in Human-Computer Interaction",
issn = "1551-3955",
publisher = "Now Publishers Inc",
number = "2-3",
}

@inproceedings{Park2020ARLooper,
	title        = {ARLooper: Collaborative Audiovisual Experience with Mobile Devices in a Shared Augmented Reality Space},
	author       = {Park, Sihwa},
	year         = 2020,
	booktitle    = {Extended Abstracts of the 2020 CHI Conference on Human Factors in Computing Systems},
	location     = {Honolulu, HI, USA},
	publisher    = {Association for Computing Machinery},
	address      = {New York, NY, USA},
	series       = {CHI EA '20},
	pages        = {1–4},
	doi          = {10.1145/3334480.3383172},
	isbn         = 9781450368193,
	numpages     = 4
}

@inproceedings{Auda2021Im,
	title        = {I’m in Control! Transferring Object Ownership Between Remote Users with Haptic Props in Virtual Reality},
	author       = {Auda, Jonas and Busse, Leon and Pfeuffer, Ken and Gruenefeld, Uwe and Rivu, Radiah and Alt, Florian and Schneegass, Stefan},
	year         = 2021,
	booktitle    = {Proceedings of the 2021 ACM Symposium on Spatial User Interaction},
	location     = {Virtual Event, USA},
	publisher    = {Association for Computing Machinery},
	address      = {New York, NY, USA},
	series       = {SUI '21},
	doi          = {10.1145/3485279.3485287},
	isbn         = 9781450390910,
	articleno    = 10,
	numpages     = 10
}

@inproceedings{Gronbaek2024Blended,
	title        = {Blended Whiteboard: Physicality and Reconfigurability in Remote Mixed Reality Collaboration},
	author       = {Gr\o{}nb\ae{}k, Jens Emil Sloth and S\'{a}nchez Esquivel, Juan and Leiva, Germ\'{a}n and Velloso, Eduardo and Gellersen, Hans and Pfeuffer, Ken},
	year         = 2024,
	booktitle    = {Proceedings of the 2024 CHI Conference on Human Factors in Computing Systems},
	location     = {Honolulu, HI, USA},
	publisher    = {Association for Computing Machinery},
	address      = {New York, NY, USA},
	series       = {CHI '24},
	doi          = {10.1145/3613904.3642293},
	isbn         = 9798400703300,
	articleno    = 798,
	numpages     = 16
}

@inproceedings{Xia2018Spacetime,
	title        = {Spacetime: Enabling Fluid Individual and Collaborative Editing in Virtual Reality},
	author       = {Xia, Haijun and Herscher, Sebastian and Perlin, Ken and Wigdor, Daniel},
	year         = 2018,
	booktitle    = {Proceedings of the 31st Annual ACM Symposium on User Interface Software and Technology},
	location     = {Berlin, Germany},
	publisher    = {Association for Computing Machinery},
	address      = {New York, NY, USA},
	series       = {UIST '18},
	pages        = {853–866},
	doi          = {10.1145/3242587.3242597},
	isbn         = 9781450359481,
	numpages     = 14
}

@inproceedings{Kunert2014Photoportals,
	title        = {Photoportals: shared references in space and time},
	author       = {Kunert, Andr\'{e} and Kulik, Alexander and Beck, Stephan and Froehlich, Bernd},
	year         = 2014,
	booktitle    = {Proceedings of the 17th ACM Conference on Computer Supported Cooperative Work \& Social Computing},
	location     = {Baltimore, Maryland, USA},
	publisher    = {Association for Computing Machinery},
	address      = {New York, NY, USA},
	series       = {CSCW '14},
	pages        = {1388–1399},
	doi          = {10.1145/2531602.2531727},
	isbn         = 9781450325400,
	numpages     = 12
}

@inproceedings{Zhang2023VRGit,
	title        = {VRGit: A Version Control System for Collaborative Content Creation in Virtual Reality},
	author       = {Zhang, Lei and Agrawal, Ashutosh and Oney, Steve and Guo, Anhong},
	year         = 2023,
	booktitle    = {Proceedings of the 2023 CHI Conference on Human Factors in Computing Systems},
	location     = {Hamburg, Germany},
	publisher    = {Association for Computing Machinery},
	address      = {New York, NY, USA},
	series       = {CHI '23},
	doi          = {10.1145/3544548.3581136},
	isbn         = 9781450394215,
	articleno    = 36,
	numpages     = 14
}

@article{Khan2024Dont,
	title        = {Don’t Block My Stuff: Fostering Personal Object Awareness in Multi-user Mixed Reality Environments},
	author       = {Khan, Talha and Lindlbauer, David},
	year         = 2024,
	month        = oct,
	journal      = {Proc. ACM Hum.-Comput. Interact.},
	publisher    = {Association for Computing Machinery},
	address      = {New York, NY, USA},
	volume       = 8,
	number       = {ISS},
	doi          = {10.1145/3698126},
	issue_date   = {December 2024},
	articleno    = 526,
	numpages     = 24
}

@inproceedings{Ouedraogo2024Where,
	title        = {Where to Draw the Line: Physical Space Partitioning and View Privacy in AR-based Co-located Collaboration for Immersive Analytics},
	author       = {Ouedraogo, Inoussa and Nguyen, Huyen and Bourdot, Patrick},
	year         = 2024,
	booktitle    = {Proceedings of the 2024 ACM Symposium on Spatial User Interaction},
	location     = {Trier, Germany},
	publisher    = {Association for Computing Machinery},
	address      = {New York, NY, USA},
	series       = {SUI '24},
	doi          = {10.1145/3677386.3682085},
	isbn         = 9798400710889,
	articleno    = 20,
	numpages     = 12
}

@inproceedings{Rasch2024Just,
	title        = {Just Undo It: Exploring Undo Mechanics in Multi-User Virtual Reality},
	author       = {Rasch, Julian and Perzl, Florian and Weiss, Yannick and M\"{u}ller, Florian},
	year         = 2024,
	booktitle    = {Proceedings of the 2024 CHI Conference on Human Factors in Computing Systems},
	location     = {Honolulu, HI, USA},
	publisher    = {Association for Computing Machinery},
	address      = {New York, NY, USA},
	series       = {CHI '24},
	doi          = {10.1145/3613904.3642864},
	isbn         = 9798400703300,
	articleno    = 952,
	numpages     = 14
}

@inproceedings{Wieland2021Separation,
	title        = {Separation, Composition, or Hybrid? – Comparing Collaborative 3D Object Manipulation Techniques for Handheld Augmented Reality},
	author       = {Wieland, Jonathan and Zagermann, Johannes and Müller, Jens and Reiterer, Harald},
	year         = 2021,
	booktitle    = {2021 IEEE International Symposium on Mixed and Augmented Reality (ISMAR)},
	volume       = {},
	number       = {},
	pages        = {403--412}
}

@inproceedings{Hubenschmid2023Colibri,
	title        = {Colibri: A Toolkit for Rapid Prototyping of Networking Across Realities},
	author       = {Hubenschmid, Sebastian and Fink, Daniel Immanuel and Zagermann, Johannes and Wieland, Jonathan and Reiterer, Harald and Feuchtner, Tiare},
	year         = 2023,
	booktitle    = {2023 IEEE International Symposium on Mixed and Augmented Reality Adjunct (ISMAR-Adjunct)},
	volume       = {},
	number       = {},
	pages        = {9--13},
	doi          = {10.1109/ISMAR-Adjunct60411.2023.00010}
}

@inproceedings{Grandi2018Design,
	title        = {Design and Assessment of a Collaborative 3D Interaction Technique for Handheld Augmented Reality},
	author       = {Grandi, Jerônimo G and Debarba, Henrique G and Bemdt, Iago and Nedel, Luciana and Maciel, Anderson},
	year         = 2018,
	booktitle    = {2018 IEEE Conference on Virtual Reality and 3D User Interfaces (VR)},
	volume       = {},
	number       = {},
	pages        = {49--56},
	doi          = {10.1109/VR.2018.8446295}
}

@inproceedings{Grandi2019Characterizing,
	title        = {Characterizing Asymmetric Collaborative Interactions in Virtual and Augmented Realities},
	author       = {Grandi, Jerônimo Gustavo and Debarba, Henrique Galvan and Maciel, Anderson},
	year         = 2019,
	booktitle    = {2019 IEEE Conference on Virtual Reality and 3D User Interfaces (VR)},
	volume       = {},
	number       = {},
	pages        = {127--135},
	doi          = {10.1109/VR.2019.8798080}
}

@article{pinho2008cooperative,
	title        = {Cooperative object manipulation in collaborative virtual environments},
	author       = {Pinho, Marcio S and Bowman, Doug A and Freitas, Carla M Dal Sasso},
	year         = 2008,
	journal      = {Journal of the Brazilian Computer Society},
	publisher    = {Springer},
	volume       = 14,
	pages        = {53--67},
	doi          = {https://doi.org/10.1007/BF03192559}
}

@inproceedings{Pintani2023CIDER,
	title        = {CIDER: Collaborative Interior Design in Extended Reality},
	author       = {Pintani, Deborah and Caputo, Ariel and Mendes, Daniel and Giachetti, Andrea},
	year         = 2023,
	booktitle    = {Proceedings of the 15th Biannual Conference of the Italian SIGCHI Chapter},
	location     = {Torino, Italy},
	publisher    = {Association for Computing Machinery},
	address      = {New York, NY, USA},
	series       = {CHItaly '23},
	doi          = {10.1145/3605390.3605419},
	isbn         = 9798400708060,
	articleno    = 16,
	numpages     = 11
}

@inproceedings{Soares2018EGO-EXO,
	title        = {EGO-EXO: A Cooperative Manipulation Technique with Automatic Viewpoint Control},
	author       = {Soares, Leonardo Pavanatto and Kopper, Regis and Pinho, Márcio Sarroglia},
	year         = 2018,
	booktitle    = {2018 20th Symposium on Virtual and Augmented Reality (SVR)},
	volume       = {},
	number       = {},
	pages        = {82--88},
	doi          = {10.1109/SVR.2018.00023}
}

@inproceedings{Pinho2002Cooperative,
	title        = {Cooperative object manipulation in immersive virtual environments: framework and techniques},
	author       = {Pinho, M\'{a}rcio S. and Bowman, Doug A. and Freitas, Carla M.D.S.},
	year         = 2002,
	booktitle    = {Proceedings of the ACM Symposium on Virtual Reality Software and Technology},
	location     = {Hong Kong, China},
	publisher    = {Association for Computing Machinery},
	address      = {New York, NY, USA},
	series       = {VRST '02},
	pages        = {171–178},
	doi          = {10.1145/585740.585769},
	isbn         = 1581135300,
	numpages     = 8
}

@article{Ruddle2002Symmetric,
	title        = {Symmetric and asymmetric action integration during cooperative object manipulation in virtual environments},
	author       = {Ruddle, Roy A. and Savage, Justin C. D. and Jones, Dylan M.},
	year         = 2002,
	month        = dec,
	journal      = {ACM Trans. Comput.-Hum. Interact.},
	publisher    = {Association for Computing Machinery},
	address      = {New York, NY, USA},
	volume       = 9,
	number       = 4,
	pages        = {285–308},
	doi          = {10.1145/586081.586084},
	issn         = {1073-0516},
	issue_date   = {December 2002},
	numpages     = 24
}

@inproceedings{Duval2006SkeweR,
	title        = {SkeweR: a 3D Interaction Technique for 2-User Collaborative Manipulation of Objects in Virtual Environments},
	author       = {Duval, T. and Lecuyer, A. and Thomas, S.},
	year         = 2006,
	booktitle    = {3D User Interfaces (3DUI'06)},
	volume       = {},
	number       = {},
	pages        = {69--72},
	doi          = {10.1109/VR.2006.119}
}

@article{benford2001collaborative,
  title={Collaborative virtual environments},
  author={Benford, Steve and Greenhalgh, Chris and Rodden, Tom and Pycock, James},
  journal={Communications of the ACM},
  volume={44},
  number={7},
  pages={79--85},
  year={2001},
  publisher={ACM New York, NY, USA}
}

@ARTICLE{Sereno2022Collaborative,
  author={Sereno, Mickael and Wang, Xiyao and Besançon, Lonni and McGuffin, Michael J. and Isenberg, Tobias},
  journal={IEEE Transactions on Visualization and Computer Graphics}, 
  title={Collaborative Work in Augmented Reality: A Survey}, 
  year={2022},
  volume={28},
  number={6},
  pages={2530-2549},
  doi={10.1109/TVCG.2020.3032761}}

@INPROCEEDINGS{Schild2018Applying,
  author={Schild, Jonas and Misztal, Sebastian and Roth, Beniamin and Flock, Leonard and Luiz, Thomas and Lerner, Dieter and Herkersdorf, Markus and Weaner, Konstantin and Neuberaer, Markus and Franke, Andreas and Kemp, Claus and Pranqhofer, Johannes and Seele, Sven and Buhler, Helmut and Herpers, Rainer},
  booktitle={2018 IEEE Conference on Virtual Reality and 3D User Interfaces (VR)}, 
  title={Applying Multi-User Virtual Reality to Collaborative Medical Training}, 
  year={2018},
  volume={},
  number={},
  pages={775-776},
  doi={10.1109/VR.2018.8446160}}

@INPROCEEDINGS{Chheang2019Collaborative,
  author={Chheang, Vuthea and Saalfeld, Patrick and Huber, Tobias and Huettl, Florentine and Kneist, Werner and Preim, Bernhard and Hansen, Christian},
  booktitle={2019 IEEE International Conference on Artificial Intelligence and Virtual Reality (AIVR)}, 
  title={Collaborative Virtual Reality for Laparoscopic Liver Surgery Training}, 
  year={2019},
  volume={},
  number={},
  pages={1-17},
  doi={10.1109/AIVR46125.2019.00011}}

@ARTICLE{Laine2024Collaborative,
  author={Laine, Teemu H. and Lee, Woohyun},
  journal={IEEE Transactions on Learning Technologies}, 
  title={Collaborative Virtual Reality in Higher Education: Students' Perceptions on Presence, Challenges, Affordances, and Potential}, 
  year={2024},
  volume={17},
  number={},
  pages={280-293},
  doi={10.1109/TLT.2023.3319628}}

@ARTICLE{Radu2023How,
  author={Radu, Iulian and Schneider, Bertrand},
  journal={IEEE Transactions on Visualization and Computer Graphics}, 
  title={How Augmented Reality (AR) Can Help and Hinder Collaborative Learning: A Study of AR in Electromagnetism Education}, 
  year={2023},
  volume={29},
  number={9},
  pages={3734-3745},
  doi={10.1109/TVCG.2022.3169980}}

@inproceedings{Li2022ARCritique,
author = {Li, Yuan and Lee, Sang Won and Bowman, Doug A. and Hicks, David and Lages, Wallace santos and Sharma, Akshay},
title = {ARCritique: Supporting Remote Design Critique of Physical Artifacts through Collaborative Augmented Reality},
year = {2022},
isbn = {9781450399487},
publisher = {Association for Computing Machinery},
address = {New York, NY, USA},
doi = {10.1145/3565970.3567700},
booktitle = {Proceedings of the 2022 ACM Symposium on Spatial User Interaction},
articleno = {10},
numpages = {12},
location = {Online, CA, USA},
series = {SUI '22}
}

@Article{Chen2021On,
AUTHOR = {Chen, Lei and Liang, Hai-Ning and Wang, Jialin and Qu, Yuanying and Yue, Yong},
TITLE = {On the Use of Large Interactive Displays to Support Collaborative Engagement and Visual Exploratory Tasks},
JOURNAL = {Sensors},
VOLUME = {21},
YEAR = {2021},
NUMBER = {24},
ARTICLE-NUMBER = {8403},
PubMedID = {34960495},
ISSN = {1424-8220},
DOI = {10.3390/s21248403}
}

@article{Yang2022Towards,
author = {Yang, Ying and Dwyer, Tim and Wybrow, Michael and Lee, Benjamin and Cordeil, Maxime and Billinghurst, Mark and Thomas, Bruce H.},
title = {Towards Immersive Collaborative Sensemaking},
year = {2022},
issue_date = {December 2022},
publisher = {Association for Computing Machinery},
address = {New York, NY, USA},
volume = {6},
number = {ISS},
doi = {10.1145/3567741},
journal = {Proc. ACM Hum.-Comput. Interact.},
month = nov,
articleno = {588},
numpages = {25}
}

@misc{tiltbrush,
  author = {Skillman \& Hackett},
  year = {2015},
  howpublished = {\url{https://support.google.com/tiltbrush/answer/6389710?hl=en}},
  title = {Tilt Brush}
}

@misc{quill,
key = {quillArt},
  howpublished = {\url{https://quill.art/}},
  title = {Quill}
}

@misc{medium,
key = {medium},
  howpublished = {\url{https://www.adobe.com/products/medium.html}},
  title = {medium by Adobe}
}

@misc{gravitysketch,
  key = {gravitysketch},
  howpublished = {\url{https://gravitysketch.com/}},
  title = {Gravity Sketch}
}

@ARTICLE{Li2003VSculpt,
  author={Li, F.W.B. and Lau, R.W.H. and Ng, F.F.C.},
  journal={IEEE Transactions on Multimedia}, 
  title={VSculpt : a distributed virtual sculpting environment for collaborative design}, 
  year={2003},
  volume={5},
  number={4},
  pages={570-580},
  doi={10.1109/TMM.2003.814795}}

@INPROCEEDINGS{Kai2006The,
  author={Kai Riege and Holtkamper, T. and Wesche, G. and Frohlich, B.},
  booktitle={3D User Interfaces (3DUI'06)}, 
  title={The Bent Pick Ray: An Extended Pointing Technique for Multi-User Interaction}, 
  year={2006},
  volume={},
  number={},
  pages={62-65},
  doi={10.1109/VR.2006.127}}

@book{jordan1996usability,
  title={Usability evaluation in industry},
  author={Jordan, Patrick W and Thomas, Bruce and McClelland, Ian Lyall and Weerdmeester, Bernard},
  year={1996},
  publisher={CRC Press}
}

@inproceedings{harms2004internal,
  title={Internal consistency and reliability of the networked minds measure of social presence},
  author={Harms, Chad and Biocca, Frank},
  booktitle={Seventh annual international workshop: Presence},
  volume={2004},
  year={2004},
  organization={Universidad Politecnica de Valencia Valencia}
}

@article{brooke1996sus,
  title={SUS: A “quick and dirty” Usability Scale},
  author={Brooke, J},
  journal={Usability Evaluation in INdustry/Taylor and Francis},
  year={1996}
}

@article{schrepp2017design,
  title={Design and evaluation of a short version of the user experience questionnaire (UEQ-S)},
  author={Schrepp, Martin and Hinderks, Andreas and Thomaschewski, J{\"o}rg},
  journal={International Journal of Interactive Multimedia and Artificial Intelligence, 4 (6), 103-108.},
  year={2017},
  publisher={UNIR}
}

@inproceedings{hart2006nasa,
  title={NASA-task load index (NASA-TLX); 20 years later},
  author={Hart, Sandra G},
  booktitle={Proceedings of the human factors and ergonomics society annual meeting},
  volume={50},
  pages={904--908},
  year={2006},
  organization={Sage publications Sage CA: Los Angeles, CA}
}

@INPROCEEDINGS{Riege2006bent,
  author={Kai Riege and Holtkamper, T. and Wesche, G. and Frohlich, B.},
  booktitle={3D User Interfaces (3DUI'06)}, 
  title={The Bent Pick Ray: An Extended Pointing Technique for Multi-User Interaction}, 
  year={2006},
  volume={},
  number={},
  pages={62-65},
  doi={10.1109/VR.2006.127}}

@inproceedings{Grandi2017Design,
author = {Grandi, Jer\^{o}nimo Gustavo and Debarba, Henrique Galvan and Nedel, Luciana and Maciel, Anderson},
title = {Design and Evaluation of a Handheld-based 3D User Interface for Collaborative Object Manipulation},
year = {2017},
isbn = {9781450346559},
publisher = {Association for Computing Machinery},
address = {New York, NY, USA},
doi = {10.1145/3025453.3025935},
booktitle = {Proceedings of the 2017 CHI Conference on Human Factors in Computing Systems},
pages = {5881–5891},
numpages = {11},
location = {Denver, Colorado, USA},
series = {CHI '17}
}

@INPROCEEDINGS{Chenechal2016Giant,
  author={Chenechal, Morgan Le and Lacoche, Jeremy and Royan, Jerome and Duval, Thierry and Gouranton, Valerie and Arnaldi, Bruno},
  booktitle={2016 IEEE Third VR International Workshop on Collaborative Virtual Environments (3DCVE)}, 
  title={When the giant meets the ant an asymmetric approach for collaborative and concurrent object manipulation in a multi-scale environment}, 
  year={2016},
  volume={},
  number={},
  pages={18-22},
  doi={10.1109/3DCVE.2016.7563562}}

@article{yu2021familiarity,
  title={Familiarity-based collaborative team recognition in academic social networks},
  author={Yu, Shuo and Xia, Feng and Zhang, Chen and Wei, Haoran and Keogh, Kathleen and Chen, Honglong},
  journal={IEEE Transactions on Computational Social Systems},
  volume={9},
  number={5},
  pages={1432--1445},
  year={2021},
  publisher={IEEE}
}

@article{janssen2009influence,
  title={Influence of group member familiarity on online collaborative learning},
  author={Janssen, Jeroen and Erkens, Gijsbert and Kirschner, Paul A and Kanselaar, Gellof},
  journal={Computers in Human Behavior},
  volume={25},
  number={1},
  pages={161--170},
  year={2009},
  publisher={Elsevier}
}

@ARTICLE{GaussianSplatting,
  author={Fei, Ben and Xu, Jingyi and Zhang, Rui and Zhou, Qingyuan and Yang, Weidong and He, Ying},
  journal={IEEE Transactions on Visualization and Computer Graphics}, 
  title={3D Gaussian Splatting as a New Era: A Survey}, 
  year={2025},
  volume={31},
  number={8},
  pages={4429-4449},
  doi={10.1109/TVCG.2024.3397828}}

@inproceedings{Greenberg1994Real,
author = {Greenberg, Saul and Marwood, David},
title = {Real time groupware as a distributed system: concurrency control and its effect on the interface},
year = {1994},
isbn = {0897916891},
publisher = {Association for Computing Machinery},
address = {New York, NY, USA},
doi = {10.1145/192844.193011},
booktitle = {Proceedings of the 1994 ACM Conference on Computer Supported Cooperative Work},
pages = {207–217},
numpages = {11},
location = {Chapel Hill, North Carolina, USA},
series = {CSCW '94}
}

@inproceedings{baron2016collaborativeconstraint,
  title={Collaborativeconstraint: Ui for collaborative 3d manipulation operations},
  author={Baron, Naem},
  booktitle={2016 IEEE Symposium on 3D User Interfaces (3DUI)},
  pages={273--274},
  year={2016},
  organization={IEEE}
}

@inproceedings{wang2021vr,
  title={VR collaborative object manipulation based on viewpoint quality},
  author={Wang, Lili and Liu, Xiaolong and Li, Xiangyu},
  booktitle={2021 IEEE International Symposium on Mixed and Augmented Reality (ISMAR)},
  pages={60--68},
  year={2021},
  organization={IEEE}
}

@article{liu2024manipulation,
  title={Manipulation guidance field for collaborative object manipulation in VR},
  author={Liu, Xiaolong and Wang, Lili and Luan, Shuai},
  journal={International Journal of Human--Computer Interaction},
  volume={40},
  number={21},
  pages={6776--6792},
  year={2024},
  publisher={Taylor \& Francis}
}
\vspace{-25pt}
\begin{IEEEbiographynophoto}{Xian Wang}
received M.Phil. degree in Artificial Intelligence at the Hong Kong University of Science and Technology. She is currently a Ph.D. candidate at the Hong Kong Polytechnic University. Her research interests include human-computer interaction, virtual reality, and collaboration.
\end{IEEEbiographynophoto}
\vspace{-25pt}
\begin{IEEEbiographynophoto}{Xuanru Cheng}
received MSc degree from the University of Glasgow. She is currently a Ph.D. student at the Hong Kong Polytechnic University. Her research interests include human-computer interaction, serious games, and virtual reality.
\end{IEEEbiographynophoto}
\vspace{-25pt}
\begin{IEEEbiographynophoto}{Rongkai Shi}
received the B.S. and Ph.D. degrees from the University of Liverpool, Liverpool, U.K., and Xi'an Jiaotong-Liverpool University, China. He is currently a postdoctoral researcher at the Hong Kong University of Science and Technology (Guangzhou), China. His research interests include human-computer interaction, extended reality, and interaction design.
\end{IEEEbiographynophoto}
\vspace{-25pt}
\begin{IEEEbiographynophoto}{Lei Chen}
received the Ph.D. degree from the University of Liverpool, Liverpool, U.K. Currently, she is a lecturer at Hebei GEO University. Her research interests include human-computer interaction, extended reality, and collaboration.
\end{IEEEbiographynophoto}
\vspace{-25pt}
\begin{IEEEbiographynophoto}{Jingyao Zheng}
received the B.S. and Master's degree in Engineering at the University of Cambridge. He is currently stuutdying at the Hong Kong Polytechnic University. His research interests include human-computer interaction, notification in VR/AR/MR, and human- centered Artificial Intelligence.
\end{IEEEbiographynophoto}
\vspace{-25pt}
\begin{IEEEbiographynophoto}{Hai-Ning Liang}
is an Associate Professor with the Computational Media and Arts Thrust at the Hong Kong University of Science and Technology (Guangzhou), China. He received his PhD in computer science from the University of Western Ontario, Canada. His research focuses on the design and evaluation of novel interactions and applications for virtual and mixed reality, gaming, and visualization technologies.
\end{IEEEbiographynophoto}
\vspace{-25pt}
\begin{IEEEbiographynophoto}{Lik-Hang Lee}
received the Ph.D. degree from the Hong Kong University of Science and Technology, and the BEng (Hons) and MPhil degrees from the University of Hong Kong. Currently, he is an assistant professor at the Hong Kong Polytechnic University. His research interests include human-centric computing systems, AR, and VR.
\end{IEEEbiographynophoto}

\vfill
\end{document}